\documentclass[aps, pra, twocolumn, groupedaddress, floatfix]{revtex4}

\usepackage{booktabs}
\usepackage{multirow}
\usepackage{siunitx}
\usepackage{amsmath}
\usepackage{amssymb}
\usepackage{graphicx}
\usepackage{float}
\usepackage[english]{babel}
\usepackage{dsfont}
\usepackage{epstopdf}
\usepackage{soul}
\usepackage{color}
\usepackage{comment}
\usepackage{braket}
\usepackage[normalem]{ulem}
\usepackage{subfiles} 
\usepackage{comment}

\usepackage{float} 

\usepackage[colorlinks=true,
            linkcolor=blue,
            urlcolor=blue,
            citecolor=blue]{hyperref}

\definecolor{SMblue}{rgb}{0,0,0}
\makeatletter
\let\SM@orig@afterheading\@afterheading
\def\@afterheading{\SM@orig@afterheading\global\everypar\expandafter{\the\everypar}}
\makeatother

\definecolor{BJorange}{RGB}{255,165,0}

\emergencystretch=\maxdimen
\begin{document}

\title{Nonlinearity-Induced Suppression of Dissipation in Bath-Coupled\\ Floquet Topological Insulators}

\author{Bhoomija Chaurasia}
\affiliation{Department of Physics, Indian Institute of Science, Bangalore 560012, India}

\author{Sebabrata~Mukherjee}
\email{mukherjee@iisc.ac.in}
\affiliation{Department of Physics, Indian Institute of Science, Bangalore 560012, India}

\date{\today}

\begin{abstract} 
Topological insulators can host chiral, disorder-robust edge states that dissipate energy when coupled to an external bath. We show that nonlinear interactions enable all-optical control over edge-state dissipation in a laser-written Floquet topological insulator periodically coupled to a photonic bath. Although moderate nonlinearity enhances dissipation due to mode mixing, increasing the nonlinear strength to an optimal value dramatically suppresses dissipation by forming highly localized traveling edge solitons. These results establish engineered  system-bath coupling and nonlinear interactions as a route to suppress dissipation of chiral topological edge transport.
\end{abstract}

\maketitle

\newpage 

\section{Introduction}
Topological physics has emerged as a vibrant field, revealing novel phases of matter that host disorder-immune boundary states~\cite{klitzing1980new, thouless1982quantized, hasan2010colloquium}.
Topological phenomena have been explored in %
various experimental settings~\cite{konig2007quantum, raghu2008analogs, wang2009observation, rechtsman2013photonic, hafezi2013imaging, aidelsburger2015measuring, susstrunk2015observation} 
that are often modeled as effectively isolated from their environment (bath).
In practice, {\color{SMblue} however, experimental systems %
can interact with %
a bath} that induces dissipation~\cite{breuer2002theory}.
Since  topological protection does not inherently extend to bath-induced dissipation, suppressing losses is crucial for enhancing the lifetime of topological states and realizing robust operation of %
topological devices 
in %
{\color{SMblue}the presence of environmental dissipation}.
However, strongly inhibiting dissipation remains challenging in purely lossy bath-coupled systems.

Controlling system-bath coupling is a versatile route for tailoring dissipation and engineering quantum states in open systems~\cite{verstraete2009quantum, harrington2022engineered}.
In particular, suitably engineered dissipation can induce topological phases~\cite{diehl2011topology} and serve as a means to realize and probe topological effects~\cite{leykam2021probing, leefmans2022topological, dreon2022self}. Nonlinear interactions provide another powerful mechanism for controlling transport beyond the linear regime~\cite{christodoulides2003discretizing, lederer2008discrete}. It is therefore natural to ask whether nonlinear interactions and engineered system-bath coupling can be combined to suppress dissipation in 
{\color{SMblue}bath-coupled} topological systems. Owing to their versatility and precise tunability, photonic devices provide an ideal platform for investigating topological phenomena in the presence of periodic driving, non-Hermiticity, and nonlinear interactions~\cite{lu2014topological, ozawa2019topological, rechtsman2013photonic, mukherjee2017experimental, maczewsky2017observation, roberts2026twisted, rudner2009topological, zeuner2015observation, mukherjee2020observation,mukherjee2021observation, maczewsky2020nonlinearity, sun2024photonic}.
{In these systems, Kerr nonlinearity arises at high optical intensities, producing an intensity-dependent local refractive index modification that can generate shape-preserving nonlinear states, known as solitons~\cite{barthelemy1985propagation, christodoulides1988discrete, eisenberg1998discrete}; see also~\cite{denschlag2000generating, trombettoni2001discrete, pernet2022gap}.}
Recent works~\cite{ablowitz2014linear, leykam2016edge, lumer2013self, mukherjee2020observation, parker2022floquet, mukherjee2021observation, maczewsky2020nonlinearity, jurgensen2021quantized, jurgensen2023quantized, dobrykh2018nonlinear, sone2024nonlinearity, tao2025nonlinearity}
have shown that nonlinear interactions can profoundly reshape topological transport, %
primarily in isolated topological systems.

Here, we realize a photonic anomalous Floquet topological insulator~\cite{rudner2013anomalous} fabricated using femtosecond laser-writing. Unlike previous works~\cite{mukherjee2017experimental, maczewsky2017observation, mukherjee2021observation}, we periodically couple the topological system to a waveguide-based photonic bath, and probe nonlinear transport at the system-bath boundary. 
{\color{SMblue} Rather than eliminating system–bath coupling, our %
approach aims to retain the bath as a functional element for selectively controlling dissipation, which 
can be useful for applications including topological lasers~\cite{bandres2018topological} and magnetic-field-free nonlinear topological isolators~\cite{zhou2017optical}.}
{\color{SMblue}We find that in} the linear regime, the system supports unidirectionally propagating edge states with lower average dissipation than the bulk states.
We show that the dissipation of these topological edge states can be either enhanced or suppressed, depending on the strength of the Kerr nonlinearity.
For an optimal nonlinear strength, a highly localized spatial soliton forms along the system-bath boundary and propagates unidirectionally, suppressing the {\color{SMblue}dissipation of light by dynamically reshaping the state to inhibit power transfer.} 
{\color{SMblue}In the presence of small defects, the soliton maintains unidirectional propagation, and the optimal power for minimum bath loss remains practically stable.}
Unlike $\mathcal{PT}$-symmetric non-Hermitian systems~\cite{bender1998real, ruter2010observation, weimann2017topologically, li2019observation, martinez2018topological, ghatak2019new,  wang2021topological, ashida2020non, miri2016nonlinearity}, where suppression of loss {\color{SMblue} usually} relies on externally supplied gain, the observed dissipation suppression arises 
in a purely lossy bath-coupled system through the interplay of nonlinearity and periodically engineered system-bath coupling.

\section{Theoretical background and model}
\begin{figure}[t]
  \centering
\includegraphics[width=1\linewidth]{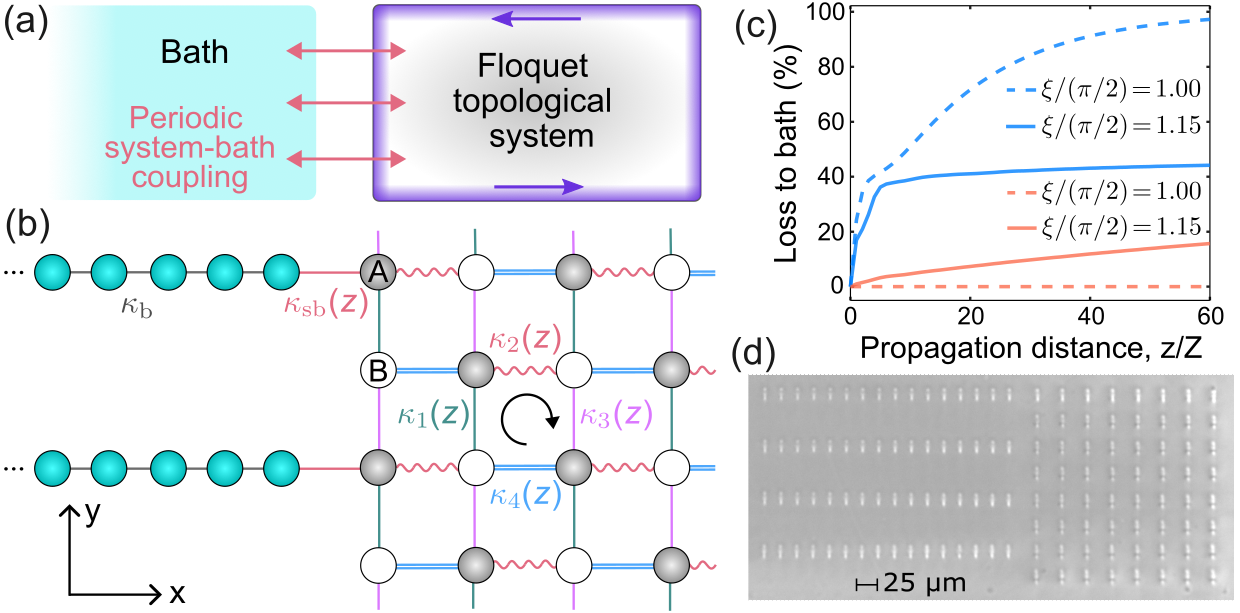}
  \caption{(a) Sketch of a topological system %
  periodically coupled to a bath on the left edge. Energy from the system can dissipate to the bath. (b) Photonic implementation of the bath-coupled topological system using waveguide networks. %
    (c) Numerically calculated power loss to the bath in the linear regime for two different system parameters: $\xi \!=\! \pi/2$ (dashed curves) and $\xi \!=\! 1.15\pi/2$ (solid curves). 
    Results are shown for excitation of the A site (red curves) and the B site (blue curves) located at the system–bath boundary. 
    (d) White-light 
    micrograph of a fs-laser-written photonic device, showing the cross-section of the system and bath waveguides.
    }
  \label{fig_1}
\end{figure}

We consider a bath-coupled photonic 
topological system [Figs.~\ref{fig_1}(a,b)] consisting of single-mode optical waveguides. The scalar-paraxial propagation of {\color{SMblue}horizontally polarized light} 
through this device {\color{SMblue}is described by} 
the following discrete nonlinear Schr{\"o}dinger-like equation~\cite{christodoulides2003discretizing, Suppmat}, equivalently, the attractive Gross-Pitaevskii equation~\cite{denschlag2000generating,trombettoni2001discrete}
\begin{eqnarray}
i\frac{\partial}{\partial z} \psi_j(z) = \sum_{\langle j' \rangle} h_{j j'} \ \psi_{j'} - g |\psi_j|^2 \psi_{j} \; , \label{eq1}
\end{eqnarray}
where
$|\psi_j|^2$ is the optical power at the $j$-th waveguide, $z$ is the propagation distance, and $h_{jj'}$ are the elements of the Hamiltonian describing the linear evolution. 
{\color{SMblue}The index $j$ runs over all waveguides 
in the system and bath.}
The second-order nonlinear effects are absent in our photonic devices created in a centrosymmetric glass material, and the
nonlinear term in Eq.~\eqref{eq1} arises due to the focusing Kerr effect.
{\color{SMblue} 
The nonlinear coefficient $g = k_0 n_2/ A_{\rm eff} > 0$, where $n_2$ is the nonlinear refractive-index coefficient, $A_{\rm eff}$ is the effective mode area, and $k_0$ is the free-space wavenumber. 
At} low power
(i.e., $gP \! \rightarrow \! 0$, where the power is $P\!=\!\sum_j |\psi_j|^2$), 
{\color{SMblue} the contribution of the nonlinear term} can be neglected. %

The topological system in Fig.~\ref{fig_1}(b) consists of a finite square lattice with a two-site basis comprising A and B sub-lattices within each unit cell. 
The nearest-neighbor couplings $\kappa_{1-4}(z)$ are periodically switched on and off in a four-step clockwise manner over a period $Z$, so that the $\kappa_m$ coupling is active during the $m$-th quarter of the driving period ($m\!=\!1,2,3,4$).
{The couplings are activated one at a time for a duration $Z/4$, so that each site connects only to one of its nearest neighbors.} 
{\color{SMblue}In the linear regime, the coupling parameter $\xi_m \!=\! \kappa_m \cdot Z/4$ determines the fraction of optical power transferred to the neighboring site, given by $\sin^2(\xi_m)$.}
For homogeneous driving $\xi_m\!\equiv\!\xi$, %
{\color{SMblue} and near the full-transfer limit, $\xi=\pi/2$,} the system is in the anomalous topological phase supporting a weakly dispersive Chern-zero bulk quasienergy band along with chiral topological {\color{SMblue}edge states; see Supplementary Section~\ref{sec:background}.} %

\begin{figure}[] 
  \centering
\includegraphics[width=1\linewidth]{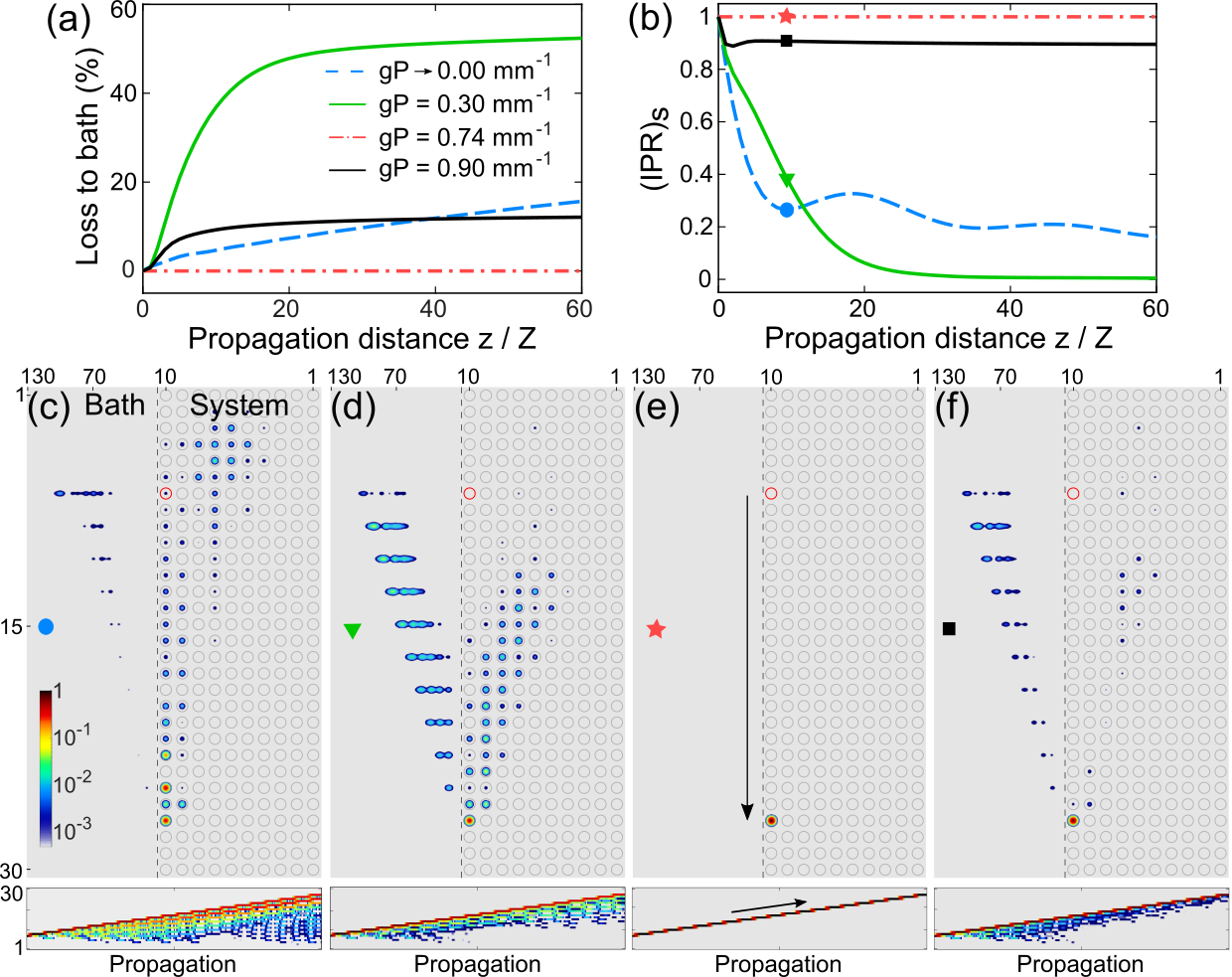}
  \caption{(a,b) Numerically obtained %
  {\color{SMblue}bath loss and $\mathrm{(IPR)_s}$ %
  versus} $z/Z$ for four %
  nonlinear strengths. (c--f) Intensity distributions at $z/Z\!=\!10$ for the four different $gP$ associated with (a,b); see the markers. The field of view %
  is smaller than the full system, and the bath is horizontally compressed for presentation purposes. 
  The initial state is launched at the boundary A site, %
  indicated by a red circle. The driving parameter $\xi$ is $1.15\pi/2$. 
  Bottom insets in (c–f) show the associated $z$-evolution of intensity at the edge waveguides on the system-bath boundary. A dissipation-less topological edge soliton is observed in (e). See also Supplementary Movies 1-4. 
}
  \label{fig_2}
\end{figure}

An engineered bath is coupled periodically to the left edge of the topological system in the following way. A semi-infinite one-dimensional array is coupled to each edge A site, and the system-bath coupling $\kappa_{sb}(z)$ is activated only during the second quarter of the driving period, i.e., when $\kappa_2$ is switched on. 
The intra-bath coupling $\kappa_b$ is %
$z$-independent.
The optical energy from the system  
can dissipate to the bath solely in the second step through the A sites. 
{\color{SMblue}This photonic bath %
 gives rise to} %
unusual phenomena, as discussed below.

{\color{SMblue} Around the full-transfer limit, the edge (bulk) modes} of the {\it isolated} system can be efficiently excited by launching light into a single A (B) site on the left/right boundary in the linear regime. Specifically, 
for $\xi \!=\! \pi/2$, an initial state localized at a single A (B) site has $100\%$ overlap with the edge (bulk) modes.
In the bath-coupled system, with $\xi \!=\! \pi/2$, light launched at an edge A site at the system-bath boundary 
propagates unidirectionally without %
dissipation %
[red dashed curve, Fig.~\ref{fig_1}(c)]. This happens simply because no light is present at the A sites during the second quarter of each driving period when the system-bath coupling is active.
Conversely, a B site excitation at the system-bath boundary  
causes significant dissipation to the bath, as the state remains localized within the system and loses power during each second quarter of the driving period. 
For $\xi \! \gtrsim \!\pi/2$,  
the edge A as well as B site excitations show dissipative propagation. However, the loss to the bath remains significantly smaller for the A-site {\color{SMblue}excitation [solid red and blue curves, Fig.~\ref{fig_1}(c)].}

The bath chains in Fig.~\ref{fig_1}(b) effectively induce loss on the A sites along the left edge, which can be captured by an imaginary on-site potential $i\gamma$ at those boundary A sites. Indeed, the dissipative dynamics of the bath-coupled model can be qualitatively captured by a non-Hermitian Hamiltonian with lossy A sites on the left edge; see Supplementary Section~\ref{H_eff_nh}. 
The quasienergy spectrum of the effective %
model hosts lossy chiral edge states, which, on average, experience less dissipation than the bulk states.
The system parameter $\xi$ provides a means to control the dissipation of initial states (see also Supplementary Section~\ref{LossVsLambda}). In a device where $\xi$ and other parameters are fixed by design, optical nonlinearity can enable all-optical control of dissipation in edge transport.

\section{Dissipation control via nonlinearity}
To investigate how the nonlinearity influences the evolution of the optical states, we 
solve Eq.~(\ref{eq1}) for different values of $gP$. %
We define the inverse participation ratio (IPR) for the state in the system as $\mathrm{(IPR)_s} = {\sum_n |\psi_n|^4}/{\big( \sum_n |\psi_n|^2 \big)^2}$
where $n$ runs over the system sites only. %
{\color{SMblue}It quantifies the localization of the state within the system 
and equals unity when} %
all the light is localized at a single site. 
The variation of loss to the bath and $(\mathrm{IPR})_s$ %
with $z/Z$ is shown in Figs.~\ref{fig_2}(a,b) for four different nonlinearities, respectively. The associated intensity distributions at a fixed %
$z/Z\!=\!10$ are shown in Figs.~\ref{fig_2}(c-f); see also Supplementary Section~\ref{method-numerics}. 
The insets in Figs.~\ref{fig_2}(c-f) show the associated intensity evolution only at the edge sites of the system close to the bath. 

In the linear regime, $gP\! \rightarrow \!0$, an initial state localized at an A site continuously dissipates light to the bath, and simultaneously spreads in the bulk of the topological lattice; Fig.~\ref{fig_2}(c). %
As a result, loss to the bath increases continuously, while %
the $\mathrm{(IPR)_s}$ first decreases rapidly and subsequently oscillates around a low baseline. %
At a moderate nonlinearity $gP \!=\! 0.30$~mm$^{-1}$, we observe %
a significantly higher loss to the bath and a rapid decay of the $\mathrm{(IPR)_s}$ toward zero as shown by the green curves in Figs.~\ref{fig_2}(a,b). The corresponding propagation dynamics are presented in Fig.~\ref{fig_2}(d); see also Supplementary Movie 2. Here, nonlinearity %
causes significant mode-mixing, resulting in the enhanced bulk diffraction and increased bath loss; see Supplementary Section~\ref{modemixing}.
Note that the %
{\color{SMblue}dissipation rate decreases at longer propagation distances as the light %
spreads into the bulk, %
moving away from the system–bath boundary
thereby} reducing the loss rate.

With increasing nonlinearity, the loss to the bath varies non-monotonically. %
After the initial increase caused by nonlinear mode-mixing, the loss decreases and vanishes at an optimal value of $gP$; see Fig.~\ref{fig_2}(e) and Supplementary Movie 3. 
At this optimal nonlinear strength, $\mathrm{(IPR)_s}$ at the end of each driving period reaches unity due to the formation of a single-site topological edge soliton. As indicated by the red lines in Figs.~\ref{fig_2}(a,b), the soliton radiates neither to the bath nor to the bulk of the lattice, and it travels one unit cell per period. 
This nonlinear bath-coupled model can also support multi-site topological edge solitons as discussed in Supplementary Section~\ref{multisite}. The edge solitons in our case are distinct from the soliton-like state reported in Ref.~\cite{mukherjee2021observation}, which slowly radiates power during propagation. It also differs from the soliton family described in Ref.~\cite{ablowitz2014linear}, which appears in the continuum limit at relatively low powers.

The {\color{SMblue}soliton formation %
can be understood from %
nonlinear power transfer between neighboring waveguides.}
For a linear coupler with $\xi \gtrsim \pi/2$, a %
fraction $\cos^2 (\xi)$ of a single-site input state returns %
to the initial waveguide after complete tunneling. %
Nonlinearity increases the propagation length required for %
complete power transfer %
[Supplementary Section~\ref{NL_coupler}].
{\color{SMblue} 
Since the soliton confines the power to at most two waveguides at the system–bath boundary [Fig.~\ref{fig_2}(e)], complete transfer over one quarter of the driving period gives
$K(\beta^2)\!=\!\xi\!=\!\kappa \cdot Z/4,$
where $\beta=gP_o/(4\kappa)$ and $K$ is the complete elliptic integral of the first kind [Supplementary Section~\ref{optimal-gp-derivation}]. At this optimal nonlinearity $gP_o$, complete power transfer suppresses both bath loss and spreading into the bulk [Fig.~\ref{fig_2}(e)].} %
{\color{SMblue}At a %
nonlinear strength of $gP=0.90$~mm$^{-1}$, the bath loss initially increases because the input state carries more power than required for soliton formation [Fig.~\ref{fig_2}(f)]. As the excess power is radiated into the bath and diffracted into the bulk, the state gradually relaxes toward the optimal soliton during propagation, resulting in weak dissipation and a nearly constant $\mathrm{(IPR)_s}$ at long propagation distances [black lines in Figs.~\ref{fig_2}(a,b)]; see 
Fig.~\ref{supp_fig_D}(c). 
The gradual relaxation of the input state toward a soliton is observed for $gP_o\!<\!gP\!<\!gP_c$, where $gP_c$ is the critical power above which bath loss increases substantially and no traveling edge soliton forms [Supplementary Section~\ref{IPRLoss}]. 
}

{\color{SMblue}%
Small defects do not cause back-scattering from the traveling soliton, and it maintains lower bath loss and stronger, more localized transmission over extended propagation distances compared to linear edge transport in the same bath-coupled system [Supplementary Section~\ref{defects} and Fig.~\ref{supp_defect_sweep}]. Furthermore, the optimal power associated with the loss minimum exhibits only small variations under weak disorder, indicating its practical stability against fabrication imperfections.
} 

{\color{SMblue}Although we %
consider a homogeneous driving parameter $\xi_m\!=\!\xi$ in Fig.~\ref{fig_2}, 
a similar %
bath} loss suppression due to the formation of an edge soliton can be achieved as long as $\xi_{1}\!=\!\xi_{3}\!\gtrsim \!\pi/2$. %
We emphasize that, for $\xi_{1,3}<\pi/2$, the edge soliton does not form, and the dissipation to the bath increases with $gP$.

\begin{figure}[] 
  \centering
  \includegraphics[width=1\linewidth]{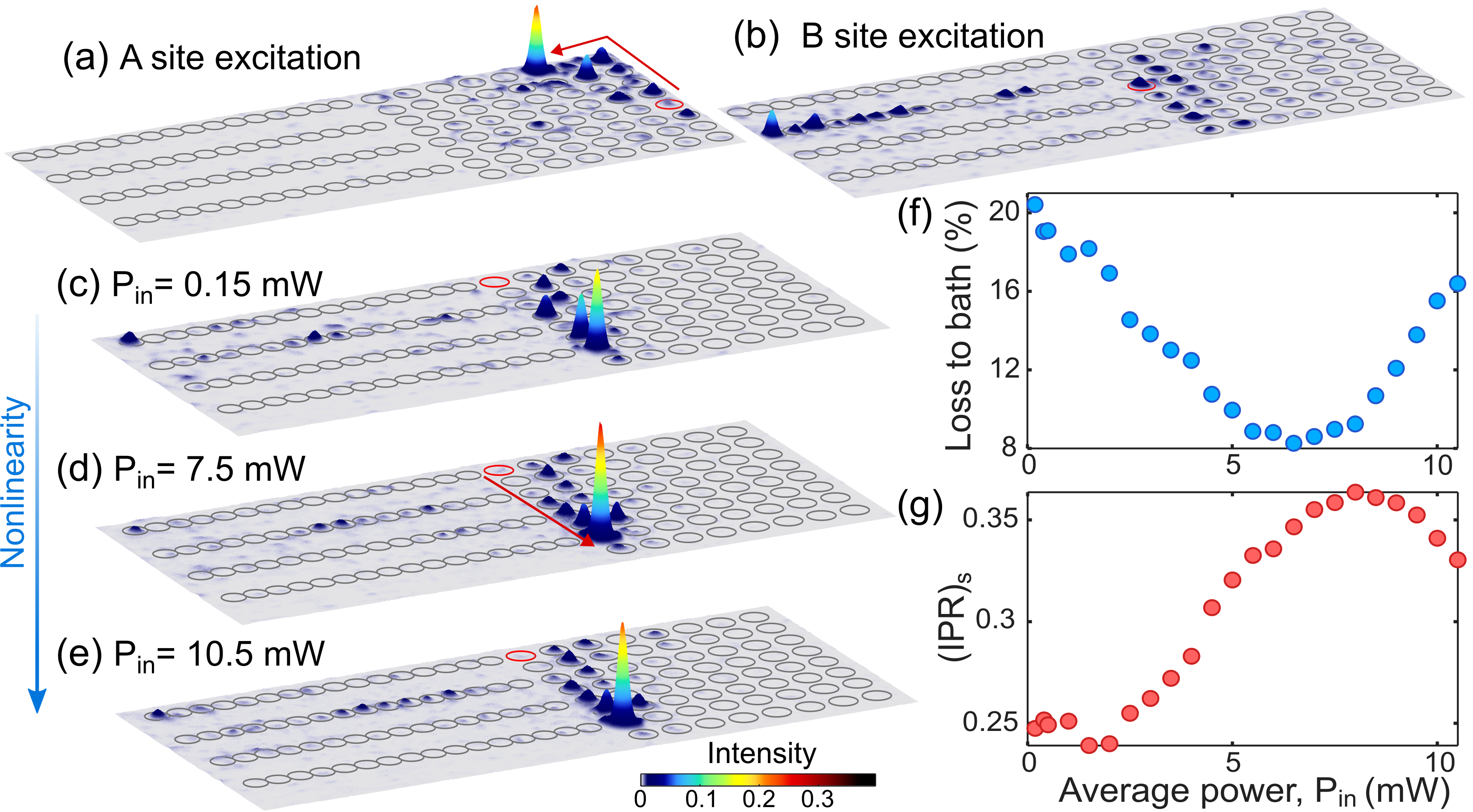}
  \caption{(a-c) Measured output intensity distributions in the linear regime for three 
  input locations. 
  (a) Counter-clockwise topological edge states on the right edge. 
  (b) Light is launched at a B site at the system-bath boundary, exhibiting a large dissipation to the bath. (c-e) A site excitation at the system-bath boundary for three different input powers, as indicated on each image.
  (f,g) For the A site excitation, 
  {\color{SMblue}the bath loss shows a dip and (IPR)$_s$ a peak as the input power is varied.}
  A highly localized unidirectional nonlinear edge state with minimal bath loss is observed in (d). All measurements are performed at $z/Z\!=\!3$ in a device with estimated 
  $\xi_{1,3} \!=\! (1.22 \pm 0.05)\pi/2 $,  $ \xi_{2,4} \!=\! (0.91 \pm 0.06)\pi/2$.  
  Red circles mark the launch sites, and gray circles denote 
  the locations of system and bath waveguides. Each intensity distribution is normalized to $1$. 
  }
  \label{fig_3}
\end{figure}

\section{Experimental implementation}
We created bath-coupled topological lattices with 
three driving periods inside a $76.2$-mm-long borosilicate glass material using femtosecond laser-writing~\cite{davis1996writing}; see Supplementary
Section~\ref{fab}. %
Each one-dimensional array in the bath contains $16$ sites, and the square lattice comprises $64$ sites.
For the maximum propagation distance considered in our experiments, the bath is long enough to ensure that light reflected from its far end does not re-enter the system. A white-light transmission micrograph of the input facet (cross-section) of the device is shown in Fig.~\ref{fig_1}(d).
All waveguides support an elliptical fundamental mode at $1030$~nm wavelength, with $1/e^2$ intensity diameters of $8.4$ and $11.5$~$\mu$m along the $x$ and $y$ axes, respectively.
The step-like couplings are implemented by bending waveguide paths --- bringing a waveguide pair into close proximity to switch on a coupling, keeping the waveguides parallel over a length $L\!=\!4$~mm, and then separating them to switch it off. Due to the elliptical mode shape, we realize a bipartite $\xi$, i.e., $\xi_1\!=\!\xi_3$ and $\xi_2\!=\!\xi_4$; 
{\color{SMblue}see Supplementary Sections~\ref{fab} and~\ref{est_lam} for} more details. 
Figures~\ref{fig_3}(a-c) show the output intensity distributions measured at $z/Z\!=\!3$ for low-power (linear) input states launched at three different locations, indicated by the red circles. In the case of Fig.~\ref{fig_3}(a), the initial state at 
the right edge of the system %
propagates counter-clockwise along the edge without any bath dissipation, as the state remains away from the bath sites. This observation clearly highlights the topological nature of the system. For Fig.~\ref{fig_3}(c), the state launched at an A site on the left edge propagates unidirectionally along the system--bath boundary with approximately $20\%$ loss to the bath.
In contrast, the input state at a B site in Fig.~\ref{fig_3}(b) leads to a significantly higher dissipation of $\approx\!49\%$, as predicted. Evidently, these linear experiments demonstrate that the topological edge states exhibit longer lifetimes compared to the bulk states. 
This dissipation contrast may be exploited in waveguide-based topological insulator lasers~\cite{bandres2018topological} to preferentially support edge-state lasing.

\begin{figure}[]
  \centering
\includegraphics[width=1\linewidth]{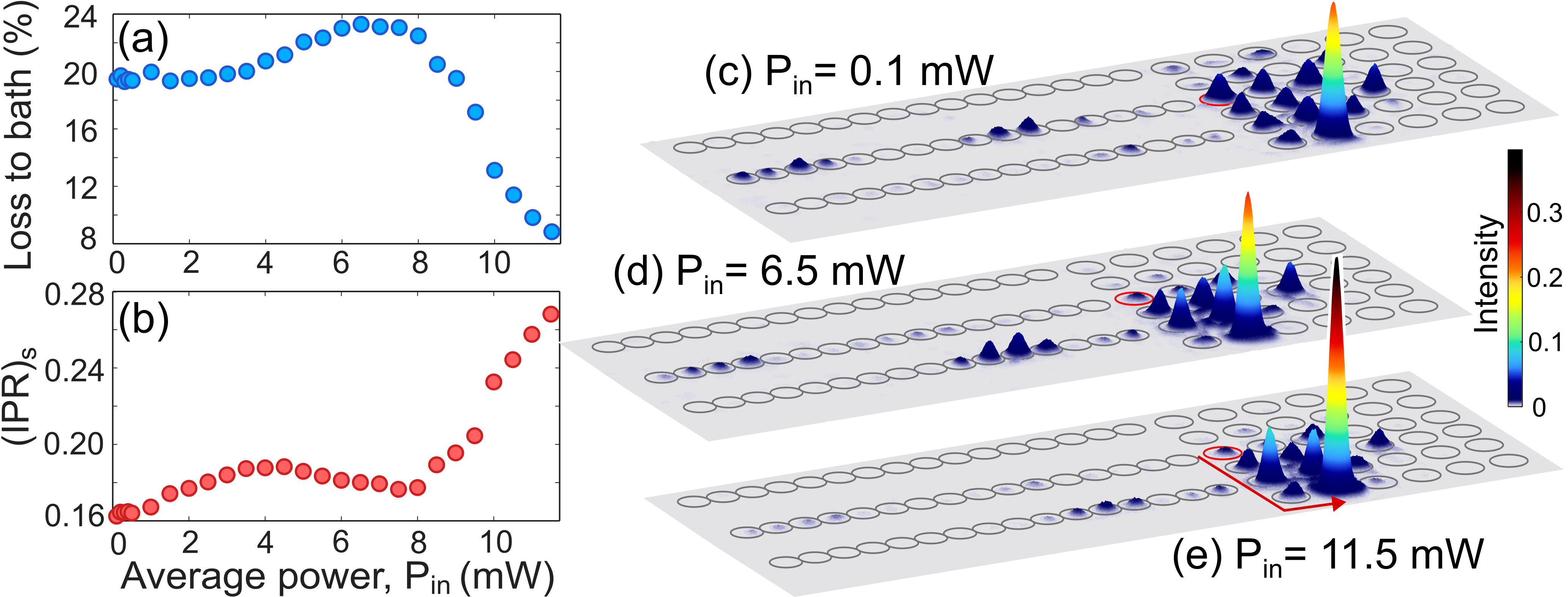}
  \caption{Observation of dissipation enhancement due to nonlinear mode-mixing.
  (a,b) Variation of loss to the bath and $(\mathrm{IPR})_s$ with $P_{\text{in}}$. Dissipation enhancement is observed in the range of $5$ to $7$~mW input power.
  (c-e) Associated experimental output intensity distributions 
  for three different $P_{\text{in}}$.
  For all measurements, light is coupled at the red-circled A site,
  and the output intensity pattern is measured at $z/Z\!=\!2$.
  }
  \label{fig_mode-mixing}
\end{figure}

We perform nonlinear experiments using laser pulses with tunable intensity and a pulse duration of $1$ ps, ensuring that temporal spreading is negligible over $z_{\text{max}}=76.2$~mm.
Figures~\ref{fig_3}(f,g) show the variation of %
bath loss and $\mathrm{(IPR)_s}$, respectively, %
as a function of the average laser power $P_{\mathrm{in}}$ for A site excitation at the system-bath boundary. %
The corresponding output intensity patterns at three different $P_{\mathrm{in}}$ values are shown in Figs.~\ref{fig_3}(c-e).
The diffraction into the lattice and dissipation into the bath observed in the linear regime [Fig.~\ref{fig_3}(c)] are significantly suppressed in Fig.~\ref{fig_3}(d).
The peak of $\mathrm{(IPR)_s}$ and the dip in the variation of bath loss both appear around $P_{\mathrm{in}}\!=\!7.5$~mW. %
The IPR peak, together with the output intensity pattern, clearly indicates the formation of a traveling edge soliton, {\color{SMblue} leading to the suppression of bath loss.} %
A similar reduction in bath dissipation and IPR peak was observed when launching light at other A sites connected to the bath, and also in measurements after a propagation of two driving periods $z/Z \!=\! 2$.

The dissipation enhancement due to nonlinear mode mixing is not evident in  Fig.~\ref{fig_3}(f). %
Within the experimentally achievable propagation distances and nonlinear strengths, %
the dissipation enhancement can be observed with relatively long straight section in the couplings. 
{We realize such a bath-coupled system with two driving periods and a comparatively smaller system size.}
In this device, the extended straight section of $L\!=\!6.2$~mm increases $\xi$ values and keeps the system-bath coupling active over a longer distance during the second driving step.
Figures~\ref{fig_mode-mixing}(a,b) show the measured loss to the bath and $\mathrm{(IPR)_s}$, respectively, 
with  $\xi_{1,3} \!=\! (1.40 \pm 0.04)\pi/2 $ and $ \xi_{2,4} \!=\! (0.95 \pm 0.09)\pi/2$. 
A clear increase in the loss to the bath is observed in the range of $5$ to $7$~mW input power.
We note that a higher nonlinearity is required to form the edge soliton at a larger $\xi$. As a result, compared to Fig.~\ref{fig_3}, the dip in bath loss and the peak in $\mathrm{(IPR)_s}$ are shifted towards higher nonlinearity. In Fig.~\ref{fig_mode-mixing}(b), the initial portion of the IPR peak is observed ($P_{\text{in}} \!=\! 7.5$ mW onwards), along with a decrease in bath-induced loss, as the maximum input power is constrained by the laser-induced damage threshold. 
The intensity distributions in Figs.~\ref{fig_mode-mixing}(c-e) clearly show dissipation enhancement at $P_{\text{in}}\!=\!6.5$~mW and a smaller dissipation at $P_{\text{in}}\!=\!11.5$~mW.

{\color{SMblue}In our experiments,} 
the loss to the bath does not  
reduce to zero %
due to intrinsic waveguide losses and unavoidable small disorder. 
Additionally, although the coupling strength decays exponentially with waveguide spacing, long-range couplings cannot be completely eliminated. Specifically, a weak system-bath coupling remains active at all $z$-values, leading to excess {\color{SMblue}linear} loss to the bath. 
{\color{SMblue}The contribution of these effects is presented in Supplementary
Section~\ref{comparison_theory_expt}, showing reasonably good agreement between theory and experiment.}

{\color{SMblue} Finally, although a specific engineered system–bath coupling is considered here, the nonlinearity-induced reduction in dissipation also occurs for a more general coupling, 
in which bath chains are coupled to both A and B sites along the system boundary throughout all four driving steps [Supplementary Section~\ref{alwayson}].
We note that random disorder appearing in experiments and the above-mentioned general system–bath coupling can limit the soliton lifetime [Supplementary Sections~\ref{defects} and \ref{alwayson}]; nevertheless, the reduction in dissipation persists over extended propagation distances.}

\section{Discussion}
In summary, we predict and experimentally demonstrate nonlinearity-induced dissipation suppression of topological edge states in a bath-coupled Floquet topological insulator. 
The interplay of nonlinearity and periodically engineered system-bath coupling gives rise to a conservative chiral edge soliton in an otherwise lossy system.
Our results establish nonlinear interactions as a tool for protecting topological states from bath-induced dissipation.
This dissipation-suppression mechanism may be extended to other bosonic platforms governed by the discrete nonlinear Schr{\"o}dinger equation~\cite{mitchell2021floquet, trombettoni2001discrete, pernet2022gap}.
%
This work raises intriguing questions about the broader role of nonlinear interactions in controlling dissipation and topology {\color{SMblue}in 
bath-coupled} systems.

\begin{acknowledgments}
We thank Archit Gajera, Trideb Shit and Gayathry Rajeevan for helpful discussions. We sincerely thank Nicholas Smith of Corning Inc.~for providing high-quality glass wafers used in the experiments. S.M.~gratefully acknowledges support from the Ministry of Education, Government of India, through the STARS program (grant number: MoE-STARS/STARS-2/2023-0716) and the Indian Institute of Science (IISc) through a start-up grant.
B.C.~thanks Council of Scientific \& Industrial Research (CSIR) for a Ph.D.~scholarship.
\end{acknowledgments}

   \vspace{12pt}
\section*{Data availability}
The data supporting the findings of this study are included within the article and its supplementary information.


%


 \bigskip
\clearpage

\clearpage


\newcommand{\beginsupplement}{%
        \setcounter{equation}{0}
        \renewcommand{\theequation}{S\arabic{equation}}%
        \renewcommand{\theHequation}{S\arabic{equation}}%
        \setcounter{figure}{0}
        \renewcommand{\thefigure}{S\arabic{figure}}%
        \renewcommand{\theHfigure}{S\arabic{figure}}%
         \setcounter{section}{0}
         \renewcommand*{\thesection}{\arabic{section}}%
         \renewcommand{\theHsection}{S\arabic{section}}%
         }
 \beginsupplement
 

\clearpage
\newpage

\onecolumngrid

\begin{center}
{\large\bfseries Supplementary Material:\par}
\vspace{3mm}
Bhoomija Chaurasia and Sebabrata Mukherjee\\[1mm]
{\small\itshape Department of Physics, Indian Institute of Science, Bangalore 560012, India\par}
\vspace{6mm}
\end{center}

\twocolumngrid

\pagenumbering{arabic} 

{\color{SMblue}
\section{Isolated topological system: spectrum and modal distributions}\label{sec:background}
}

\begin{figure}[]
  \centering
  \includegraphics[width=0.95\linewidth]{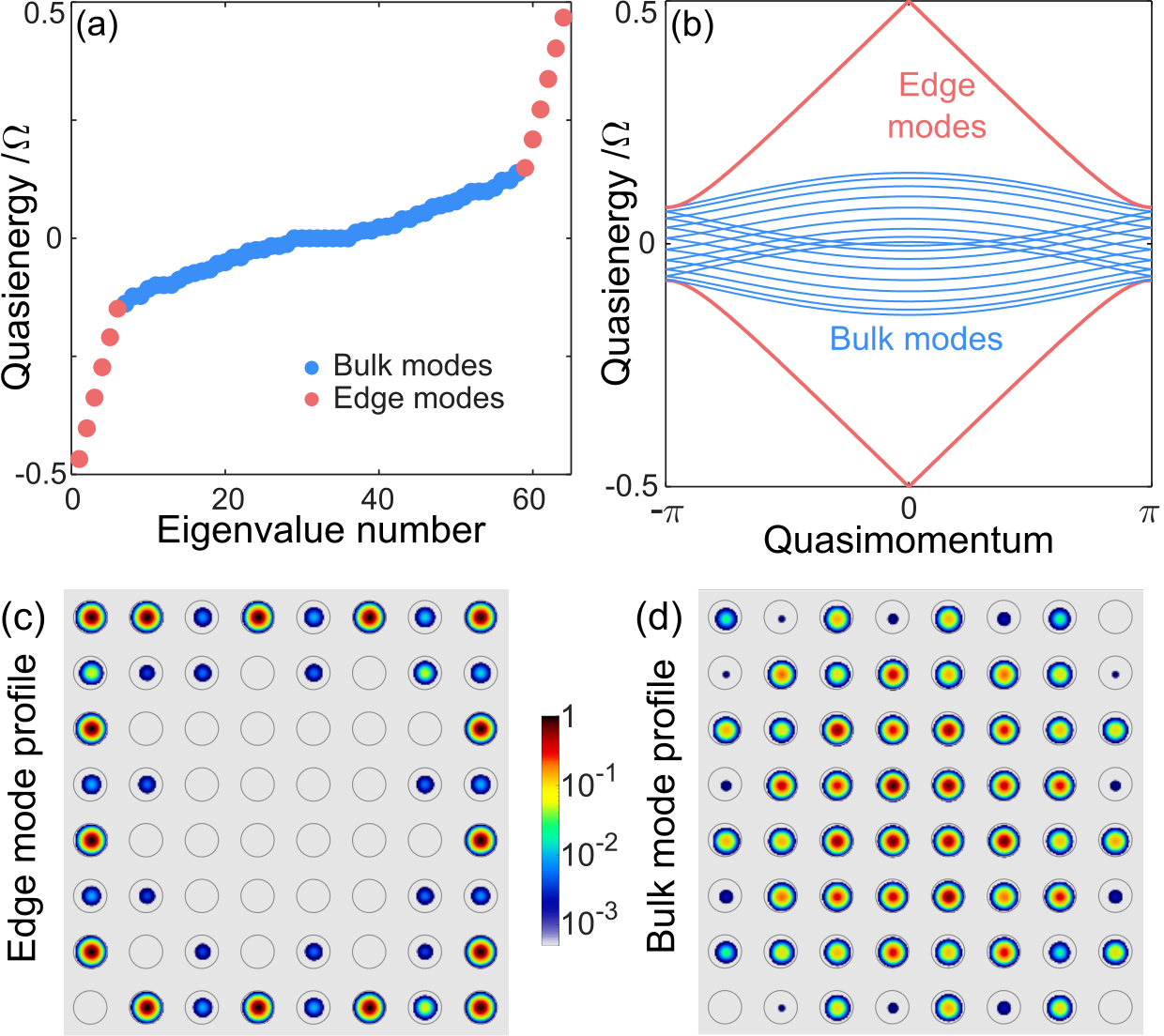}
  \caption{{\color{SMblue}
  (a) Quasienergy spectrum of an $8\times8$ %
  isolated topological system without the bath. %
  Here, $\xi=1.15\pi/2$. 
  (b) The dispersion of the bulk and edge modes is shown considering a strip geometry periodic along the vertical axis.
  Intensity profiles of an edge mode at $\varepsilon/\Omega=0.47$ and a bulk mode at $\varepsilon/\Omega=0.14$ are shown in (c) and (d), respectively. %
  The colormap is in a logarithmic scale.} 
  }
  \label{S1}
\end{figure}

{\color{SMblue} In this section, we present the Floquet quasienergy spectrum and the modal distributions of a bath-free, isolated topological system. We first consider a finite-size 
$8\times8$ %
square lattice with the four-step driving protocol, as mentioned in the main text. The quasienergy spectrum is calculated by diagonalizing the Floquet operator over one driving period. Figure~\ref{S1}(a) shows the spectrum for  $\xi=1.15\pi/2$, where the bulk and edge modes are marked in blue and red.

The dispersion of the bulk and edge modes is shown in Fig.~\ref{S1}(b), where the quasienergy spectrum is plotted considering a strip geometry, periodic along the vertical and finite along the horizontal axis. Note that the system supports a single bulk band, whose Chern number is therefore zero. Nevertheless, chiral edge modes traverse the quasienergy gap, demonstrating anomalous Floquet topology~\cite{rudner2013anomalous}, in which chiral edge states can exist despite a vanishing bulk Chern number. The topological character of this system is captured by the winding number, which is unity in this case. 
}

{\color{SMblue} Figures~\ref{S1}(c,d) present the real-space distribution of an edge and a bulk mode, respectively. Notice that the edge mode is localized on the boundaries with an exponential decay into the bulk. 
Additionally, at the left edge (where the bath is coupled in Fig.~\ref{fig_1}), all edge modes have significantly higher weight on the A sites, enabling their efficient excitation %
by launching light into an A site at that edge.
The bulk modes are completely delocalized and have nearly equal weight on both A and B sites in the entire lattice.
During the second quarter of the driving, the total projection of edge (bulk) states on the B (A) sites of the left boundary is higher than that on the A (B) sites. 
Since the bath is coupled to the boundary A sites during the second quarter, this contrast in the projection provides an intuitive explanation for the larger linear dissipation for the B site excitation compared to the A site excitation [Fig.~\ref{fig_3}(b,c)].
}

\section{Effective non-Hermitian model}\label{H_eff_nh}

\begin{figure}[t]
  \centering
  \includegraphics[width=\columnwidth]{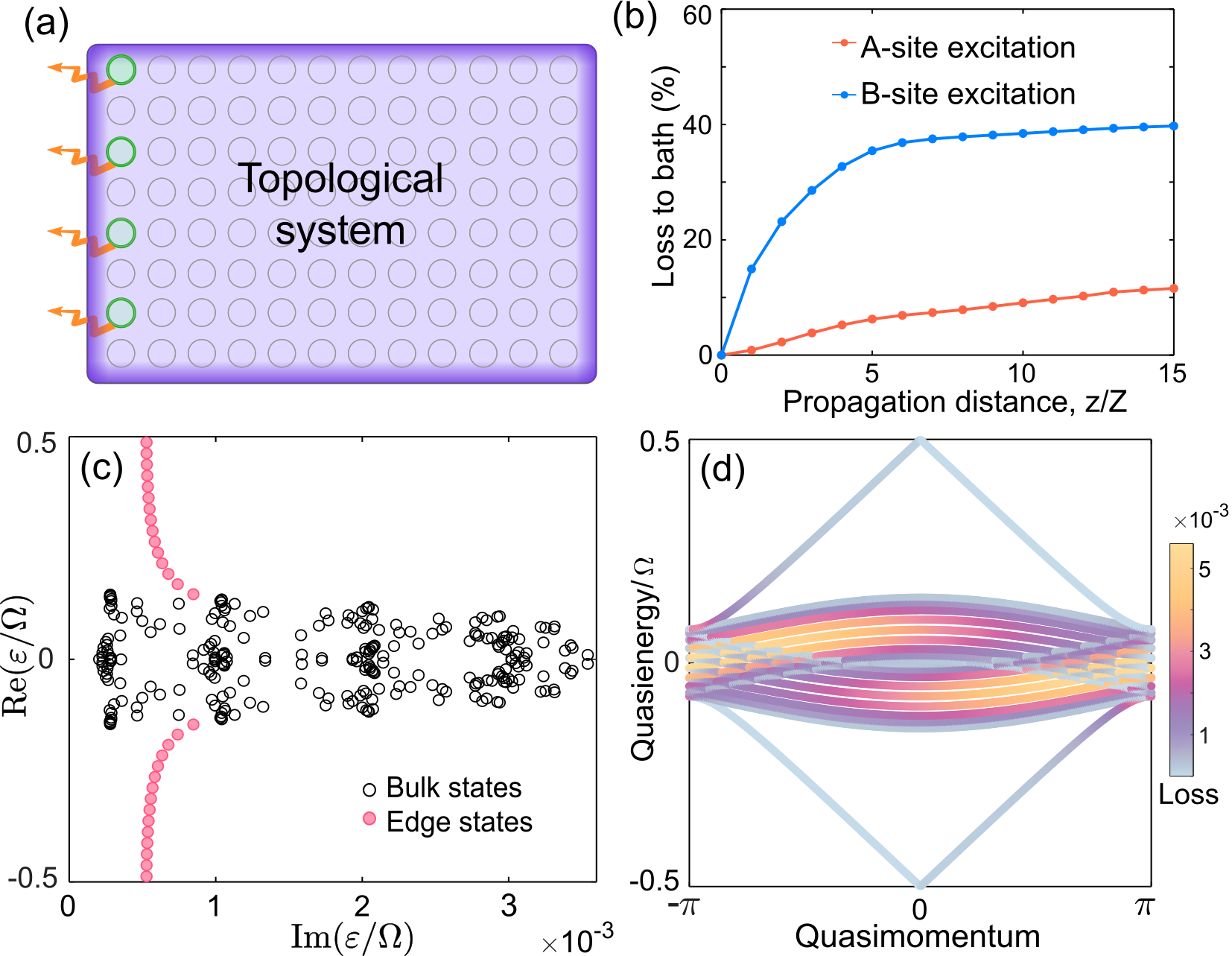}
  \caption{(a) An effective non-Hermitian description of Fig.~\ref{fig_1}(b). The effect of the bath chains is qualitatively captured by an imaginary on-site energy at the boundary A sites, active only during the second quarter of the driving period.
  (b) Loss to the bath vs $z/Z$, for A site and B site excitations at the lossy boundary, qualitatively agrees with Fig.~\ref{fig_1}(c). Here, $\xi\!=\!1.15\,\pi/2$ and $\gamma\!=\!0.0312$~mm$^{-1}$. 
  (c) Quasienergy spectrum,  %
  showing the real vs imaginary part of the eigenvalues. On average, the edge modes are less lossy compared to the bulk modes.
  (d) Quasienergy spectrum considering a strip geometry periodic along the $y$ direction. Here, the color map indicates the  $\text{loss = -Im}(\varepsilon / \Omega)$.
}
  \label{supp_figC}
\end{figure}

In this section, we discuss how the quasienergy spectrum is obtained for a bath-coupled topological system.
To this end, we present an effective non-Hermitian description of the bath-coupled topological system. Note that the bath chains in Fig.~\ref{fig_1}(b) effectively induce loss on the A sites along the left edge of the topological system. This bath-induced loss can be captured by an imaginary on-site potential $i\gamma$ at the boundary A sites; see Fig.~\ref{supp_figC}(a). In our case, this imaginary on-site potential is introduced only during the second quarter of the driving period, when the system-bath coupling $\kappa_{sb}$ is active in the bath-coupled model.
We determine the loss coefficient 
$\gamma$ by optimizing the agreement between the $z$-evolution of power loss to the bath for the bath-coupled model and the effective non-Hermitian model.
The variation of the loss for single A and B site excitations on the lossy boundary is shown in Fig.~\ref{supp_figC}(b); %
{\color{SMblue} see also the solid curves} in Fig.~\ref{fig_1}(c). 
Dissipation is substantially larger for initial excitation at the edge B site than for excitation at the A site. 
{\color{SMblue}As mentioned in the main text,} for B site excitation, the loss rate to the bath decreases after some propagation, when the state spreads into the bulk of the system, moving away from the left edge.

Using the effective non-Hermitian Hamiltonian, we compute the complex quasienergies, which show the key differences between
edge and bulk modes. In Fig.~\ref{supp_figC}(c), the quasienergies
highlighted in magenta are edge modes with smaller
average imaginary parts than bulk modes, indicating
reduced loss for the edge modes. In the $  k  $-space spectrum in Fig.~\ref{supp_figC}(d), we show the real quasienergy with loss
visualized as a colormap. This figure shows that the
chiral edge modes have lower loss, particularly when
the quasienergy approaches the 
$\pi$-gap. This effective
non-Hermitian model is considered only in this section
to compute the spectrum. For numerical calculations
presented in other sections, we have considered the model
described in Fig.~\ref{fig_1}(b).

\section{Loss vs $\xi$}\label{LossVsLambda}

\begin{figure}[]
  \centering
  \includegraphics[width=0.9\columnwidth]{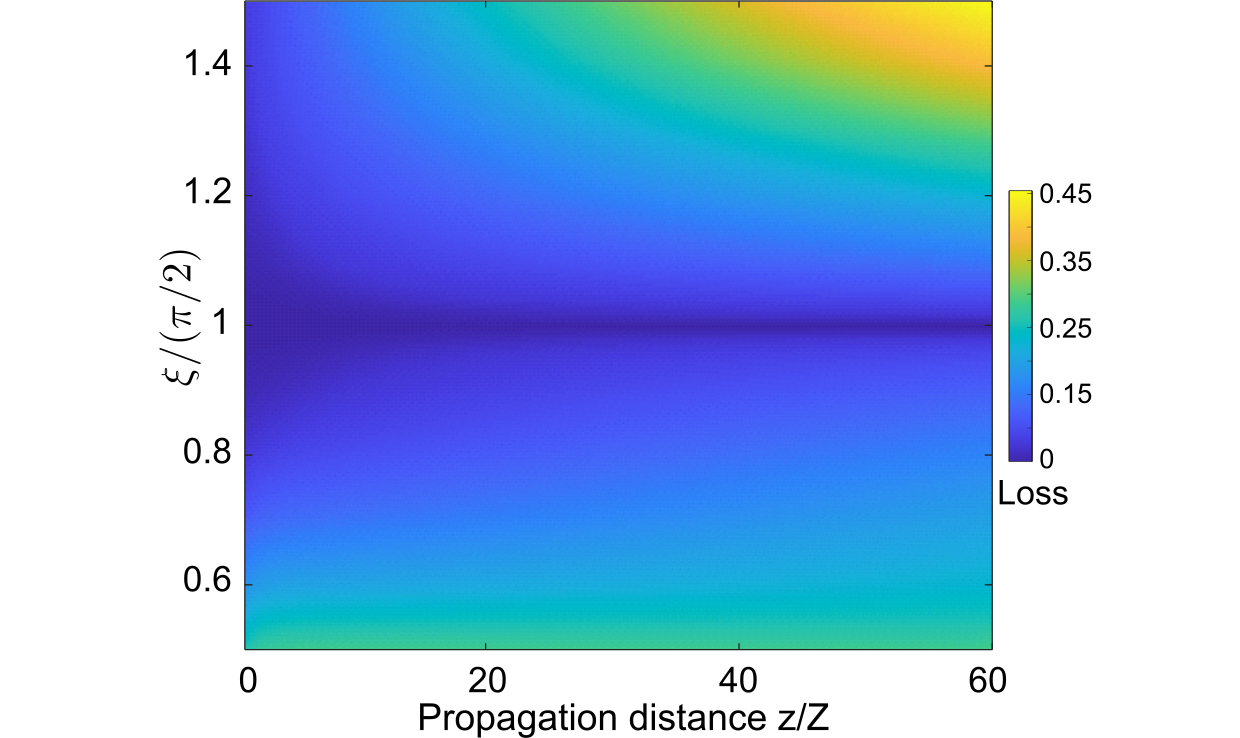}
  \caption{%
  Bath loss as a function of driving period $z/Z$ and $\xi$. Loss vanishes at the $\xi\!=\!\pi/2$ and grows asymmetrically as $\xi$ deviates from $\pi/2$. %
}
  \label{supp_fig3}
\end{figure}

In this section, we discuss how the dissipation of light into the bath varies as a function of the system parameter $\xi$ in the linear regime.
We consider system-bath and intra-bath couplings of $\kappa_{sb}\!=\!0.15$ mm$^{-1}$ and $\kappa_b\!=\!0.19$~mm$^{-1}$, respectively. The couplings in the system $\kappa_{1-4}$ are varied to tune $\xi_m\!=\!\xi$ from $\pi/4$ to $3\pi/4$, and this particular range of $\xi$ is chosen because the system is topological within this interval.
Figure~\ref{supp_fig3} shows the variation of optical power loss to the bath as a function of $\xi$ and propagation distance $z/Z$ for a single-site initial state at
an A site on the system-bath boundary. Here, the power loss at a given $z$ is defined as $\mathcal{L}\!=\!1-\sum_n |\psi_n(z)|^2$, where the index $n$ runs over the system sites only.
The dissipation of topological edge states vanishes at $\xi \!=\! \pi/2$ and increases asymmetrically around this $\xi$ value. The lattice model and the coupling protocol to the bath enable the dissipation tunability by controlling $\xi$.  In Fig.~\ref{supp_fig3}, for $ \tilde{\xi} \equiv \xi/(\pi/2) \geq 1$, the power loss $  \mathcal{L}$ as a function of $\tilde{z}\equiv z/Z$ can be well described by the following empirical expression
\begin{eqnarray}
\mathcal{L}(\tilde{\xi},\tilde{z}) = A(\tilde{\xi}) \, e^{-B(\tilde{\xi})\, \tilde{z}} + C(\tilde{\xi}) \, ,
\end{eqnarray}
where $A$, $B$, and $C$ are polynomial functions of $ \tilde{\xi}$. %
The coefficients of the empirical polynomials are given in Table~\ref{empirical_fit}.

\begin{table}[t]
\centering
\caption{
Fitted coefficients of the empirical polynomials}
\label{empirical_fit}
\vspace{14pt}
\begin{tabular}{|c|c|c|c|}
\hline
Polynomial & Degree & Coefficient & Value \\
\hline
\multirow{5}{*}{$A$} & \multirow{5}{*}{4}
    & $a_4$ & 58.3724767 \\
    & & $a_3$ & -280.648701 \\
    & & $a_2$ & 503.400004 \\
    & & $a_1$ & -400.907412 \\
    & & $a_0$ & 119.783632 \\
\hline
\multirow{3}{*}{$B$} & \multirow{3}{*}{2}
    & $b_2$ & 0.144933806 \\
    & & $b_1$ & -0.376009768 \\
    & & $b_0$ & 0.255204134  \\
\hline
\multirow{5}{*}{$C$} & \multirow{5}{*}{4}
    & $c_4$ & -45.4393095 \\
    & & $c_3$ & 216.395762 \\
    & & $c_2$ & -384.630755 \\
    & & $c_1$ & 304.182575 \\
    & & $c_0$ & -90.5082727 \\
\hline
\end{tabular}
\end{table}

As mentioned in the main text, $\xi$ and other parameters are fixed by design for a given photonic device. In this situation, optical nonlinearity is a valuable tool for all-optical dissipation control for the edge transport (see also Section~\ref{IPRLoss} and Fig.~\ref{supp_fig_D}%
). 

{\color{SMblue}
\section{Experimental details}\label{supp_expt}
}

\subsection{Design and fabrication of photonic devices}\label{fab}
\begin{figure}[]
  \centering
  \includegraphics[width=0.9\columnwidth]{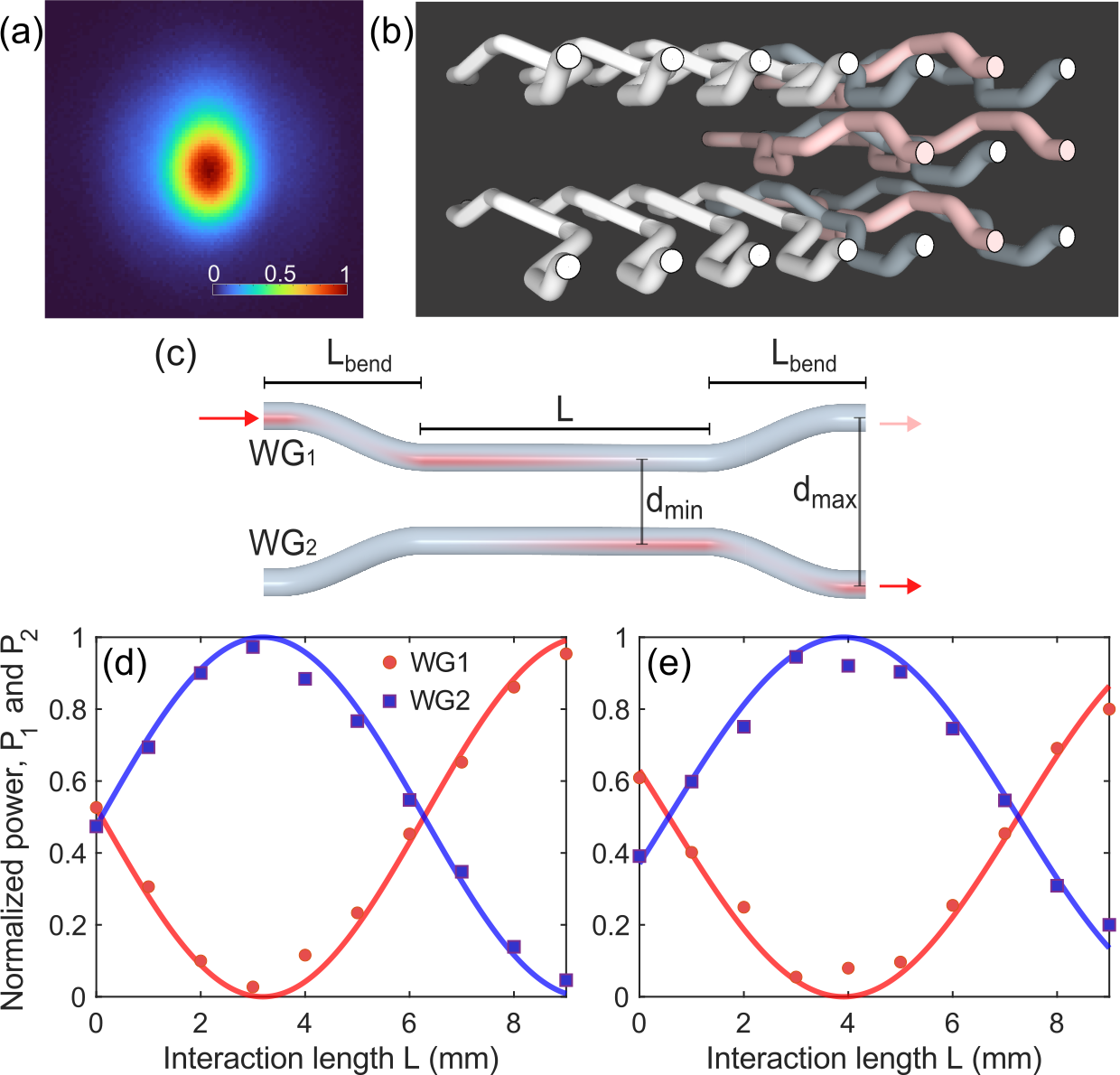}
  \caption{
{\color{SMblue}(a) Measured mode profile at $1030$~nm, with mode-field diameters of 8.4~$\mu$m and 11.5~$\mu$m along the horizontal and vertical directions, respectively. (b) Simplified schematic of a $3\times 3$ topological system with two bath chains (light gray) on the left. (c) Schematic of two synchronously bent waveguides showing how a coupling is switched on and off by adjusting the inter-waveguide spacing. (d,e) Experimentally measured and numerically obtained normalized power in waveguide 1 and 2 as a function of $L$ for vertical and horizontal couplings.}
}
  \label{hc_vc}
\end{figure}

Photonic lattices are fabricated using fs laser writing inside a 76.2-mm-long Eagle-XG glass material. Fabrication parameters were optimized to realize low-loss, single-mode waveguides at $1030$~nm wavelength of light. %
All waveguides in the device are fabricated with the same set of parameters, such as laser pulse energy, polarization, and the translation speed of fabrication. {\color{SMblue}Figure~\ref{hc_vc}(a) shows the measured mode profile at $1030$~nm. In this case, the}
mode field diameters ($1/e^{2}$ intensity) 
are $\mathrm{MFD}_{x} \! = \! 8.4 \pm 0.1~\mu$m and $\mathrm{MFD}_{y} \! = \! 11.5 \pm 0.1 ~\mu$m in the horizontal and vertical directions, respectively.

As mentioned in the main text, the topological lattices are realized using a four-step Floquet driving protocol, to which a photonic bath is periodically coupled along one edge. 
{\color{SMblue}Figure~\ref{hc_vc}(b) shows a simplified sketch of a $3\times 3$ topological system with two bath chains (light gray color) on the left side.} 
The lattice in Fig.~\ref{fig_3} was fabricated over a depth range of $100-282\, \mu$m below the top surface of the glass material. 
{\color{SMblue}Couplings in the device are switched on and off by adjusting the spacing between adjacent waveguides -- Fig.~\ref{hc_vc}(c) shows an example.}
The A and B waveguides of the lattice are initially separated by {\color{SMblue}$d_{\text{max}}=\,$}26$~\mu$m.
In each Floquet step, every pair of waveguides is synchronously bent to reduce their separation and thereby switch on the coupling. The inter-waveguide spacing is then kept fixed 
{\color{SMblue}at $d_{\text{min}}$ over} 
a desired length $L$ 
{\color{SMblue}after which} 
waveguides are bent apart in a reverse manner to switch off the coupling. 
For the results shown in Figs.~\ref{fig_3} and \ref{fig_mode-mixing}, the minimum separations are 14.8~$\mu$m for the horizontal couplings ($\kappa_{2,4}$) and 15.8~$\mu$m for the vertical couplings ($  \kappa_{1,3}  $). %
In Fig.~\ref{fig_3}, bend sections of $1.5$~mm and straight sections of length $L\!=\!4$~mm are used, whereas in Fig.~\ref{fig_mode-mixing}, bend sections of $2$~mm and straight sections of $6.2$~mm are used.
The bath comprises arrays of waveguide chains. Each chain is connected to an edge A site of the topological lattice only during the second step of the Floquet period. This is achieved by {\color{SMblue}bending the bath waveguide towards the lattice during the second quarter of the driving step,} reducing the system–bath separation from 29~$\mu$m to 18.8~$\mu$m.
{\color{SMblue} A fixed coupling between the bath waveguides is realized by
maintaining a constant intra-bath separation of 16.8~$\mu$m for all $z$.} 
Note that the edge A sites are coupled to two sites during the second step. %

\subsection{Loss estimation}
{\color{SMblue}We perform insertion loss (propagation loss + bend loss + coupling loss) measurements using free-space input/output coupling. 
The coupling loss arises primarily due to 
the Fresnel reflection from both facets of the glass material and the loss due to mode mismatch. The propagation loss was estimated by characterizing straight optical waveguides and was found} to be {$0.45$~dB/cm}. 
The system and bath waveguides were designed to have the same number of bends and similar bend profiles, thereby ensuring the same overall intrinsic losses. 
{\color{SMblue} The total bend loss due to each bend section of the topological lattice is estimated to be $\sim 0.2\,\mathrm{dB}$.}

{\color{SMblue}
\subsection{Estimation of disorder strengths}
To quantify fabrication-induced disorder, %
we fabricate and characterize two-waveguide directional couplers, as shown in Fig.~\ref{hc_vc}(c). The experimental data in Figs.~\ref{hc_vc}(d) and (e) correspond to the vertical ($\kappa_1$) and horizontal ($\kappa_2$) couplings, respectively. For each set, ten couplers are fabricated keeping the bend sections identical and varying the straight interaction length $L$. 
The minimum inter-waveguide spacing for the couplers in Fig.~\ref{hc_vc}(d) and Fig.~\ref{hc_vc}(e) are $d_{\text{min}}\!=\! \{16.6, \, 15.6\}~\mu$m, respectively. 
Horizontally polarized, low-power light at $1030$~nm is launched into waveguide 1 (WG1), and the output powers $P_1$ and $P_2$ are measured as a function of $L$. 
The normalized power $P_2/(P_1+P_2)$ at waveguide 2 exhibits a squared-sinusoidal dependence on $L$: $\frac{\kappa^2}{\kappa^2+(\Delta \beta/2)^2}\cdot \sin^2\big(\sqrt{\kappa^2+(\Delta \beta/2)^2}\, \cdot L + \phi \big)$, where $\kappa$ is the coupling, $\Delta\beta$ is the on-site propagation constant mismatch, and $\phi$ is a phase factor arising due to the bending sections; see also Sec.~\ref{NL_coupler}.

We fit the experimental data using the coupled-mode model. The solid curves in Fig.~\ref{hc_vc}(d, e) are obtained by minimizing the sum of squared errors (SSE), which gives R-squared values near $0.97$. %
The two waveguides are fabricated with identical parameters, and we observe near-complete power transfer. These suggest that the waveguides can be assumed to be identical with vanishing propagation constant mismatch. In other words, the small deviations of the experimental data points from the fitted solid line can be explained using random fluctuations in couplings. The mean and standard deviation of couplings are estimated to be
$\{\kappa, \, \delta \kappa\}\!=\!\{ 0.252, \, 0.010\}$~mm$^{-1}$ and $\{ 0.238, \, 0.014\}$~mm$^{-1}$
for Fig.~\ref{hc_vc}(d, e), respectively. 
Although the estimated $\delta \beta$ is negligible, we account for its effect on soliton propagation by introducing random on-site propagation constant disorder to the entire topological lattice  [Supplementary Section~\ref{defects}].
}

{\color{SMblue}
\subsection{Single-mode operation and polarization preservation}

{\it Single-mode operation:} To investigate whether the fabricated waveguides support higher-order spatial modes at the operating wavelength of $1030$~nm, we perform the following experiment. %
We couple near-monochromatic light filtered from a supercontinuum source into a straight isolated waveguide and image the output optical field profile as a function of wavelength. 
During each measurement, the focused beam position at the input of the waveguide is gradually shifted in the transverse directions to detect whether the output Gaussian intensity profile changes.
A higher-order spatial mode was observed near $805$~nm wavelength, whereas no higher-order mode is resolved above this cutoff wavelength; see Figs.~\ref{mode_Approx}(a,b). These observations are consistent with single-mode operation at the wavelength used in our experiments for the input conditions considered here.

\begin{figure}[]
  \centering
\includegraphics[width=0.95\linewidth]{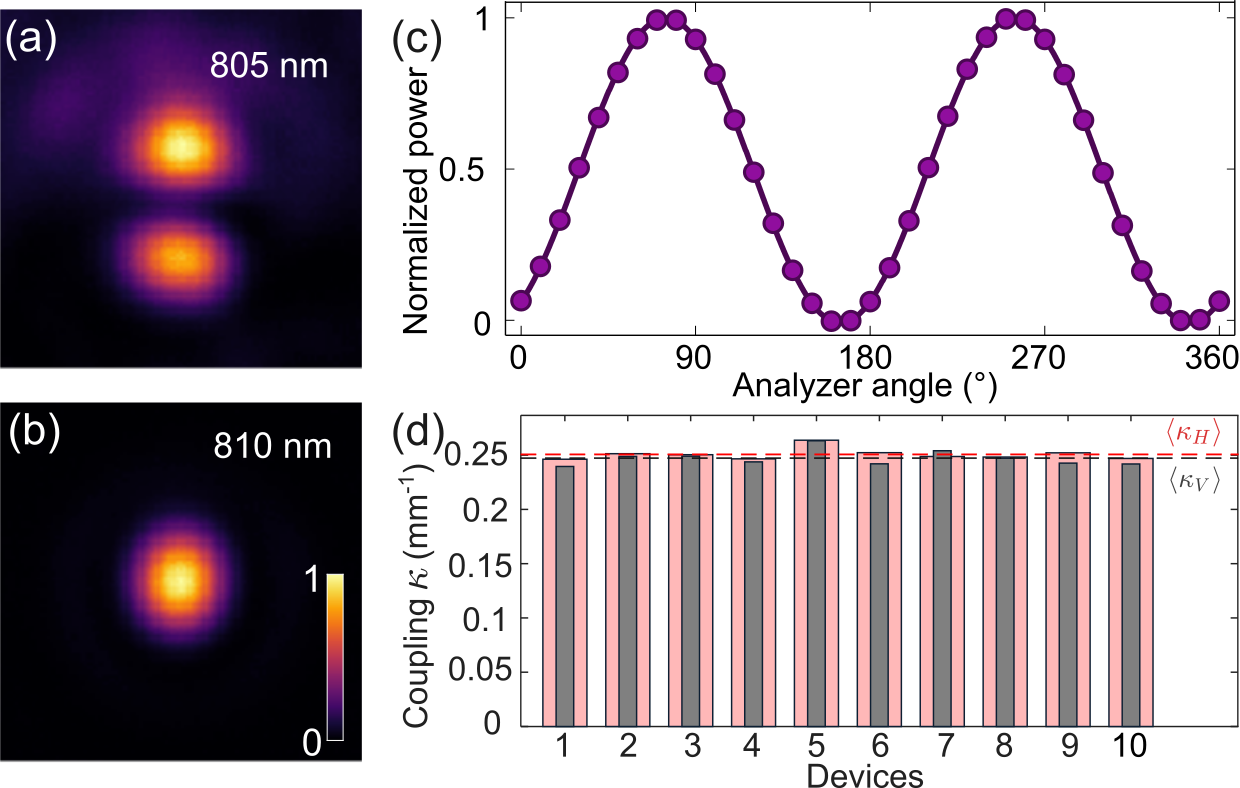}
  \caption{{\color{SMblue} (a,b)  Intensity profiles measured at the output of an isolated 
waveguide %
at wavelengths of $805$ and $810$ nm. %
  (c) Normalized transmitted power as a function of analyzer angle. Circles denote measured data, and the solid curve is a fit to Malus's law. %
  (d) The measured coupling strengths for incident horizontal and vertical polarization in $10$ different devices. The coupling for the vertical polarization is about $1 \%$ smaller.}}
  \label{mode_Approx}
\end{figure}

{\it Polarization preservation:} We perform all experiments by launching horizontally polarized light at $1030$~nm wavelength into a desired waveguide of the photonic devices.
To investigate polarization preservation over the maximum propagation distance considered in our experiments, we measure the output polarization of light after propagation through (a) an isolated waveguide with the same bend profile as those in the lattice and (b) a Floquet topological lattice. In both cases, we analyze the output using a linear polarizer and monitor the transmitted power as a function of the analyzer angle. %
The transmitted intensity decreases by more than $98.3\%$ from its maximum to its minimum, indicating that the output remains predominantly linearly polarized.
This result %
supports the use of a scalar-wave model to describe the horizontally polarized light dynamics over the propagation distances considered.

We also characterize the polarization dependence of the inter-waveguide coupling using directional couplers. The measured coupling strength for vertical polarization is on average about $1 \%$ smaller than that for horizontal polarization. %
Thus, %
our waveguide devices exhibit small polarization dependence. 
We therefore employ the scalar-wave approximation for the experimental configuration considered in this work, without assuming polarization-independent device performance.}

\begin{figure}[]
  \centering
\includegraphics[width=0.95\linewidth]{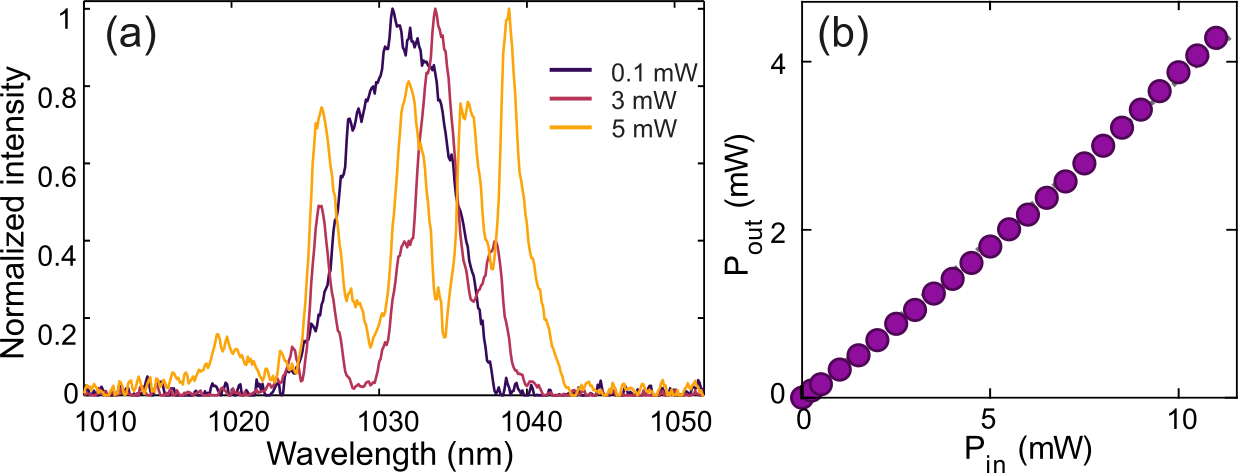}
  \caption{{\color{SMblue}(a) Variation of normalized spectral power at different average input powers. %
  (b) Measured linear variation of output power with the input power, indicating that nonlinear absorption is negligible in our experiments.}
  }
  \label{SPM}
\end{figure} 

{\color{SMblue}
\subsection{Self-phase modulation and nonlinear loss}

In our experiments, optical Kerr nonlinearity is accessed using $1$ picosecond, down-chirped laser pulses from an Amplitude Satsuma laser operating at a repetition rate of $5$~kHz. Figure~\ref{SPM}(a) shows the variation of normalized spectral power at different average input powers. At $0.1$~mW input power, the spectral width (FWHM) is $8$~nm. As the input power is increased, the spectrum broadens due to self-phase modulation (SPM). %
Since coupling strength $\kappa$ depends on wavelength, the generated spectral components can experience slightly different coupling.

The variation of $\kappa$ with wavelength $\lambda$ is experimentally measured using directional couplers and tuning the wavelength of the optical source in the range of $1000$ to $1060$~nm in steps of $5$~nm. A linear increase in $\kappa(\lambda)$ is observed in this wavelength range with $d\kappa/d\lambda = 0.30\%$~mm$^{-1}$ per nm (not shown). 
Notice that the spectral broadening due to SPM remains around $14$~nm after $76$~mm of propagation. %
Consequently, the variation in $\kappa$ across the broadened spectrum due to SPM is similar to the fabrication-induced fluctuations in the coupling strength.} 

{\color{SMblue}During all nonlinear experiments, we monitor both the input and output powers to assess potential nonlinear absorption at high intensities. As shown in Fig.~\ref{SPM}(b), the output power varies linearly with the input power over the range investigated, indicating that nonlinear absorption is negligible under our experimental conditions.
}

\section{Nonlinear coupler:\\ increase in full-transfer length}\label{NL_coupler}
In this section, we discuss a nonlinear directional coupler consisting of two waveguides that are brought into close proximity over a length $L$ and then separated {\color{SMblue}[Fig.~\ref{hc_vc}(c)]}. 
The propagation of the optical field in waveguides~1 and 2 can be governed by the following equations
\begin{eqnarray}
    i\frac{\partial}{\partial z} \psi_{1}=-i\alpha \psi_{1} -\kappa(z) \, \psi_{2} - g|\psi_{1}|^2\psi_{1} \; , \label{eqn_NL_coupler1}\\
i\frac{\partial}{\partial z} \psi_{2}= -i\alpha \psi_{2} -\kappa(z)\, \psi_{1} -g |\psi_{2}|^2\psi_{2} \; ,
\label{eqn_NL_coupler2}
\end{eqnarray}
where $\kappa(z)$ is the evanescent coupling, and $\alpha$ is the attenuation coefficient.

We fabricated a set of $20$ horizontal directional couplers, varying the length $L$ 
from $0.5$ to $10$~mm in steps of $0.5$~mm. For all these devices, the bending profiles are identical.
In each device, the minimum inter-waveguide separation is $d_{\text{min}}=15.6~\mu$m in the coupling region, the length of each bend section is $9$~mm, and the waveguides are bent apart to a separation of $d_{\text{max}}=60~\mu$m.

\begin{figure}[]
  \centering
\includegraphics[width=1\columnwidth]{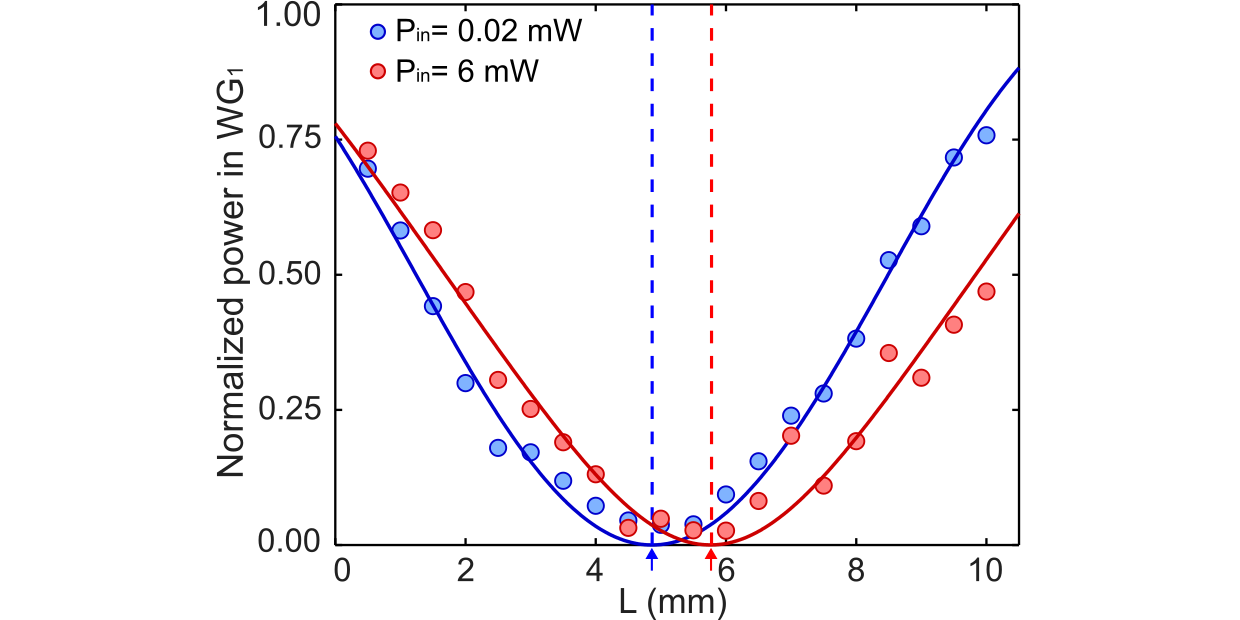}
  \caption{The variation of normalized output intensity in waveguide~1 as a function of coupling length $L$. The blue and red circles are measured data points, %
 and the solid lines are obtained numerically. 
 The blue and the red arrows indicate the length $L$ for full transfer of optical power in the linear and nonlinear cases, respectively.}
  \label{supp_fig2}
\end{figure}

When low-power ($gP \rightarrow \!0$) light is initially launched into waveguide~1, the power variations at the two sites are given by $|\psi_1(z)|^2 = e^{-2\alpha z}\cos^2(\xi)$ and $
|\psi_2(z)|^2 = e^{-2\alpha z}\sin^2(\xi)$.
Denoting the coupling strength at the straight section by $\kappa_0$, we can write $\xi\!=\! \kappa_0 L +\phi$, where $\phi$ is the contribution from the bending sections.
Evidently, the optical power, normalized at each $z$, i.e., $|\psi_{2, 1}|^2/(|\psi_1|^2+|\psi_2|^2)$, exhibits squared sinusoidal/cosinusoidal variations with $z$.

The blue circles in Fig.~\ref{supp_fig2} show the experimentally measured variation of normalized power as a function of $L$ in the linear regime. The solid blue line is obtained numerically, yielding a fitted coupling strength  $\kappa_0  \!=\!  0.217 $~mm$ ^{-1} $ and $ \phi  \!=\!  0.517$ rad.
The full-transfer length, i.e., the minimal length $L$ required for the normalized power at the waveguide~1 to reach zero, is given by $ L \!=\! 4.87 $~mm. Note that this length corresponds to  $ \xi  \!=\!  \kappa_0 L + \phi \!=\! \pi/2 $.

In the nonlinear regime ($g\! \neq \! 0$), Eqs.~\eqref{eqn_NL_coupler1},\eqref{eqn_NL_coupler2} can be analytically solved for two coupled straight waveguides in the absence of loss ($\alpha\!=\!0$)~\cite{jensen2003nonlinear}. In this case, the power at waveguide~1 varies as
$ |\psi_1(z)|^2  \!=\!  \frac{1}{2} |\psi_1(0)|^2 \left[1 + \mathrm{cn} \big(2 \kappa z + 2\phi \, \middle| \, \big(\frac{|\psi_1(0)|^2}{\mathcal{P}_c}\big)^2 \big)\right] $,
where $  \mathrm{cn}  $ is the Jacobi elliptic {cosine} function, and $\mathcal{P}_c$ is the critical power which denotes the threshold above which nonlinearity detunes the coupling, inhibiting full power transfer~\cite{jensen2003nonlinear}. %
Note that at low nonlinearity, $|\psi_1(z)|^2$ is a periodic function, exhibiting full transfer with an increased period.

For our nonlinear {\it bendy couplers}, we numerically solve
Eqs.~\eqref{eqn_NL_coupler1}, \eqref{eqn_NL_coupler2} considering the experimentally realized parameters. Measured normalized power at waveguide~1 and its numerically obtained variation with $L$ is shown in Fig.~\ref{supp_fig2}, see the red circles and the red curve. For an attenuation coefficient $\alpha \!=\! 0.0052\,\mathrm{mm}^{-1}$, experimental data agree well with the numerics for a fitted $g\!=\!0.090\,\mathrm{mm}^{-1}\mathrm{mW}^{-1}$. We perform similar experiments for four different input laser powers, varying from $2$~mW to $7.4$~mW, and estimate the nonlinear coefficient to be
$g \!=\! 0.091 \pm 0.005 \,\mathrm{mm}^{-1}\mathrm{mW}^{-1}$.

It is evident from Fig.~\ref{supp_fig2} that the length required for the full transfer of optical power increases in the nonlinear regime. For $P_{\text{in}}\!=\!6$~mW, this length is $5.67$~mm.
Within the range of optical nonlinearity achieved in our experiments, we observed this tunability of full transfer length with input power.
Note that this feature cannot be observed at larger nonlinearity, which can cause a partial transfer or trapping of optical power at waveguide~1.

These measurements confirm that high-intensity light can dynamically restore full-transfer conditions in the over-coupled regime ($\xi \gtrsim\pi/2$) through intensity-dependent refractive index modulation. In the bath-coupled topological lattice, this effect suppresses diffraction and enables the formation of propagating solitons, as discussed in the main text.

{\color{SMblue}
\section{Condition for soliton formation}\label{optimal-gp-derivation}

In this section, we discuss how the optimal nonlinear strength for soliton formation, 
$gP_o$, is related to other device parameters, such as $\xi$, $Z$ and $\kappa$. 
It is worth noting that, when the soliton forms at 
$gP_o$, the optical power is confined to at most two waveguides at the system-bath boundary. 
In this situation, the results in Section~\ref{NL_coupler} can be readily applied to the bath-coupled topological system. We first derive the condition for complete transfer of optical power from an edge A site to the B site directly below it over a propagation distance $Z/4$.

Considering single-site excitation at an edge A waveguide, Eqs.~\eqref{eqn_NL_coupler1} and \eqref{eqn_NL_coupler2} can be combined into a single equation for $\alpha\!=\!0$ and step coupling profile:
\begin{equation}
\left(\frac{dS}{dz}\right)^2
=
4\kappa^2(1-S^2)\bigl[1-\beta^2(1-S^2)\bigr]. \label{eq:Sode}
\end{equation}
Here, $S \equiv  (|\psi_A|^2-|\psi_B|^2)/{P}$ is the power imbalance and $\beta \equiv {gP_o}/(4\kappa)$. Eq.~\eqref{eq:Sode} supports the following exact solution in terms of the Jacobi elliptic cosine function:
\begin{equation}
S(z)=\operatorname{cn}(2\kappa z,\beta^2). \label{eq:cn}
\end{equation}
For a complete transfer of optical power from the A to B site, 
the function \(\operatorname{cn}(2\kappa z, \, \beta^2)\) should be $-1$, and the minimal value of $2\kappa z$ required for this is given by $\kappa z =K(\beta^2)$, where
\[
K(\beta^2)=\int_0^{\pi/2}\frac{d\theta}{\sqrt{1-\beta^2\sin^2\theta}}
\]
is the complete elliptic integral of the first kind, and $S=\cos(\theta)$.
Setting $z=Z/4$, the optimal nonlinear strength for the edge soliton formation can be written as
\begin{equation}
K(\beta^2)=\xi=\kappa \cdot Z/4 \label{eq:thresh}
\end{equation}
}
{\color{SMblue} Eq.~\eqref{eq:thresh} admits unique oscillatory (z-periodic) solutions for
$\beta\!\in\![0,1]$ only if $\xi\!>\!\pi/2$. In other words, no edge soliton forms for $\xi_{1,3}\!<\!\pi/2$, as stated in the main text.}

 \section{Numerical calculations}\label{method-numerics}
For numerical evolution of a state according to Eq.~\eqref{eq1}, we consider a topological system of size $N_x \times N_y \!=\!40 \times 130$ sites, with each bath chain containing $800$ sites.
Eq.~\eqref{eq1} is solved numerically using the  Runge–Kutta solver ode45 in MATLAB, with open boundary conditions. The field amplitudes are propagated along the $z$ coordinate for a given initial state.
It should be noted that couplings are modeled as ideal step functions with maximum values of $\kappa_{1-4}=0.28$~mm$^{-1}$, $\kappa_{sb}=0.15$~mm$^{-1}$ over $Z/4\!=\! 6.5$ mm. In comparison, when the couplings are step-like functions with smoother rise and fall profiles, as in our experiments, the same phenomena are observed, with a reduced
nonlinear strength; see Section~\ref{Expt_step_like}. %

\section{Variation of IPR and bath loss with nonlinearity}\label{IPRLoss}

\begin{figure}[h!]
  \centering
  \includegraphics[width=\columnwidth]{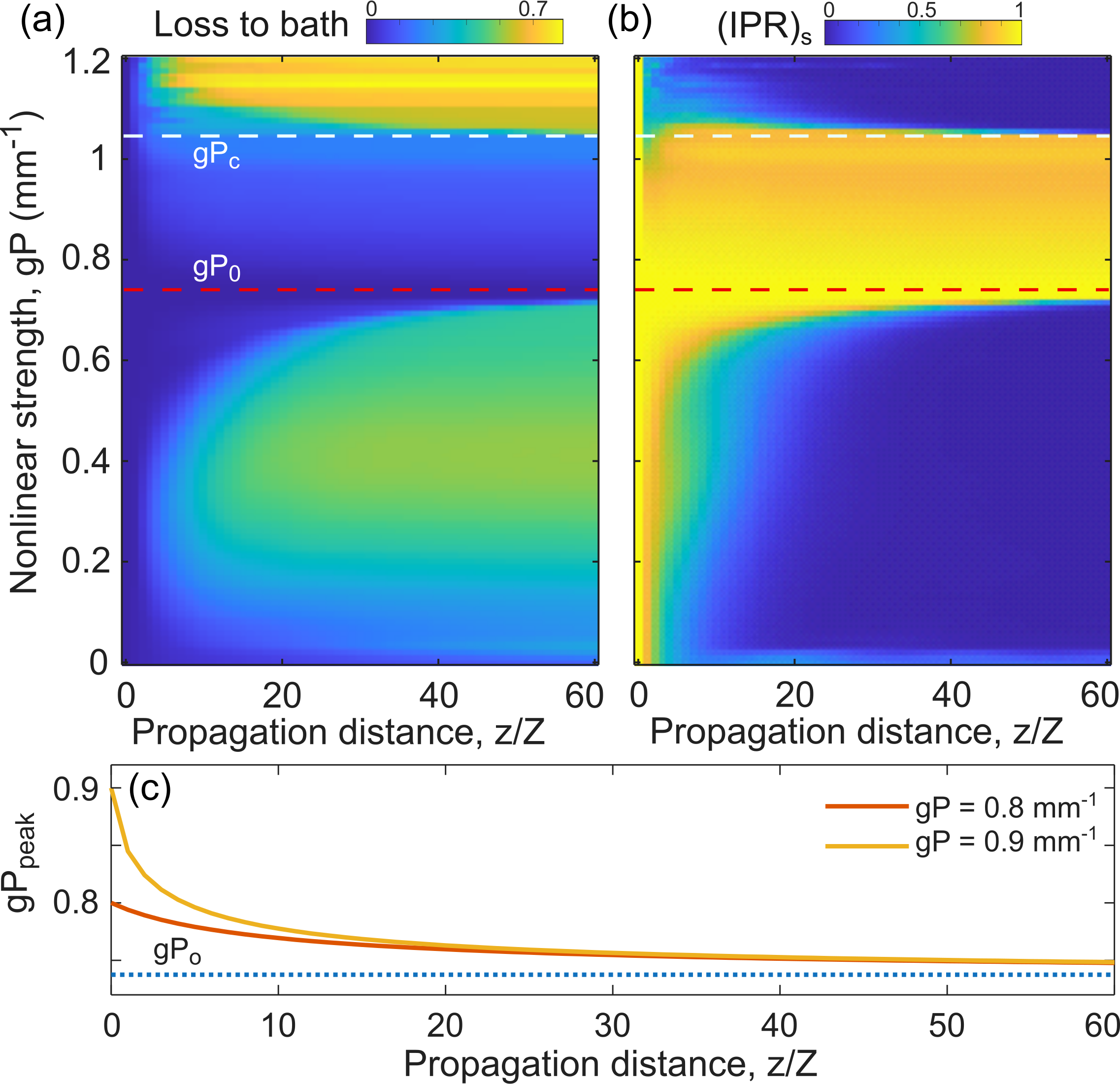}
  \caption{Numerically obtained variation of (a) the power loss to the bath and (b) $\mathrm{(IPR)_s}$ as a function of the nonlinear strength $gP$ and Floquet periods $z/Z$. The horizontal red dashed line indicates the nonlinear strength at which the soliton forms at the boundary. The white dashed line marks the critical power $gP_c=1.05$~mm$^{-1}$ above which the infinite lifetime soliton does not form. Here, $\xi\!=\!1.15 \pi/2$. {\color{SMblue}(c) The variation in optical power at the most intense site as it propagates along the boundary, showing that the nonlinear state gradually approaches the optimal value (blue dotted line). The initial $gP$ values are 0.8 and 0.9 mm$^{-1}$.}
  }
  \label{supp_fig_D}
\end{figure}

Figures~\ref{supp_fig_D}(a,b) present the variation of the loss to the bath and $\mathrm{(IPR)_s}$ for a range of nonlinear strengths {\color{SMblue}for a disorder-free lattice}. The horizontal red dashed line indicates the value of nonlinearity $gP\!=\!0.74$~mm$^{-1}$ at which the single-site soliton forms at the boundary. In this case, the loss to the bath vanishes, and the $\mathrm{(IPR)_s}$ reaches its maximum. As a function of the nonlinear strength, the $\mathrm{(IPR)_s}$ exhibits a peak, whereas the loss to the bath displays a dip, 
as confirmed experimentally in Fig.~\ref{fig_3}. 
We have considered the same system size and coupling parameters as mentioned in the main text and Section~\ref{method-numerics}. 
However, in each bath chain, we now consider $N_b \!=\! 50$ sites with absorbing boundary conditions by introducing the imaginary onsite energy at the far end of the bath. Specifically, the imaginary onsite energy increases exponentially toward the end of the bath, with a maximum loss at the end site. The propagation of initial states localized at a single A site on the system-bath boundary is computed by solving Eq.~\eqref{eq1}.

{\color{SMblue}
When $gP$ exceeds the optimal value, both the bath loss and $\mathrm{(IPR)_s}$ show small variations at large propagation distances over a finite range of nonlinear strength. Beyond this range, indicated by the white dashed line in Figs.~\ref{supp_fig_D}(a,b), both quantities deteriorate rapidly. Within the range $gP_o<gP<gP_c$, the initial nonlinear state has more power than is required for soliton formation. During propagation, the excess optical power is radiated into the bath and diffracted into the bulk of the system. Consequently, the state traveling along the system-bath boundary gradually relaxes to the soliton at long propagation distances. The optical power at the most intense site, $gP_{\text{peak}}$ [see, e.g., Fig.~\ref{fig_2}(f)], is plotted along the propagation direction in Fig.~\ref{supp_fig_D}(c). The results show that the state gradually approaches the optimal value, marked by the blue dotted line. This relaxation is algebraic rather than exponential, resulting in a slow variation of the loss and $\mathrm{(IPR)_s}$ over many periods.}

{\color{SMblue}%
For input powers above $gP_c$ [see Supplementary Movie 5] within the range considered here, the high nonlinear strength permits only partial power transfer from the edge A site to the neighboring B site, such that $gP_{\text{peak}}<gP_o$ after the first driving period. 
Additionally, substantial optical power is lost to the bath during the second driving step, 
causing the total power remaining in the system to rapidly %
fall below $gP_o$. 
The power remaining is insufficient to form the soliton at the system–bath boundary. 
As a result, relatively low $\mathrm{(IPR)_s}$ and substantial bath loss are observed, as shown in Figs.~\ref{supp_fig_D}(a,b). We obtain  $gP_c$ by finding the power value above which $gP_{\text{peak}}<gP_o$ after the first driving period.
}


\begin{figure}[t]
  \centering
  \includegraphics[width=0.90\columnwidth]{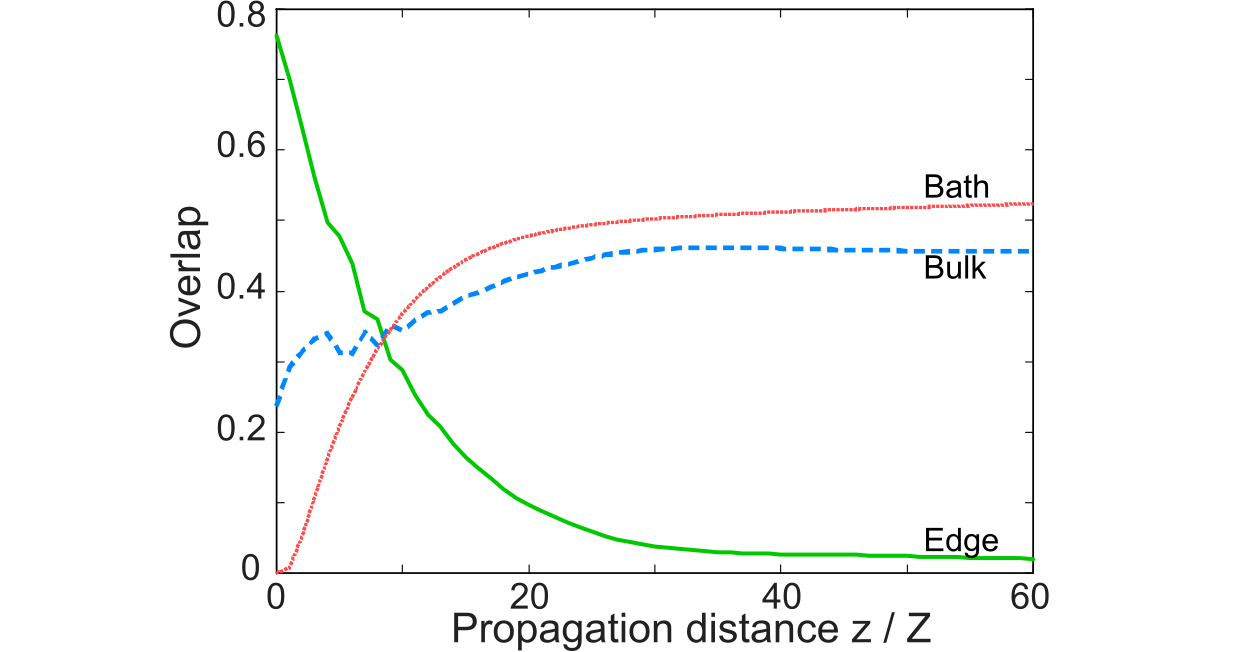}
  \caption{Nonlinear mode-mixing. The variation of the overlaps of the normalized nonlinear state $\psi_{\text{sys}}(z)$ in the system with the bulk (blue dashed) and edge (green solid) eigenstates of the isolated linear system. The 
  red line shows the bath overlap, indicating the increase in bath loss with propagation distance.
  Here, $gP \!=\! 0.30$~mm$^{-1}$ and $\xi\!=\!1.15 \pi/2$.}
  \label{supp_fig_E}
\end{figure}  

\section{Nonlinear mode-mixing}\label{modemixing}

As discussed in the main text, the bath loss significantly increases and the $\mathrm{(IPR)_s}$ rapidly decays toward zero at moderate nonlinear strengths due to nonlinear mode-mixing; see Figs.~\ref{fig_2}(a,b,d) and Supplementary Movie~2. In this section, we provide additional results to further support this claim.

We propagate the single A site input state launched at the system-bath boundary for $60$ Floquet periods with a nonlinear strength $gP \!=\! 0.30$~mm$^{-1}$. 
The system component $\psi_{\text{sys}}(z)$ of the normalized wave function is then extracted. The total overlaps of $\psi_{\text{sys}}(z)$ with the edge and bulk Floquet eigen modes $\Phi_n$ of the isolated linear system are computed as
\begin{eqnarray}
{\text{Overlap}}(z) = \sum_{n} |\langle \Phi_n | \psi_{\text{sys}}(z) \rangle|^2\, ,
\label{overlap}
\end{eqnarray}
where $n$ are the indices of the edge and bulk modes. The bath overlap is calculated by subtracting the edge and bulk overlaps from the total overlap of unity. 
Figure~\ref{supp_fig_E} displays the overlap with edge and bulk states, along with the bath contribution.
The initial state has a significant overlap with the edge states. However, as the state evolves, its overlap with the bulk and bath states grows and then saturates. This mode-mixing induced by the nonlinearity causes bulk spreading and loss to the bath. 

\begin{figure}[t]
  \centering
  \includegraphics[width=0.95\linewidth]{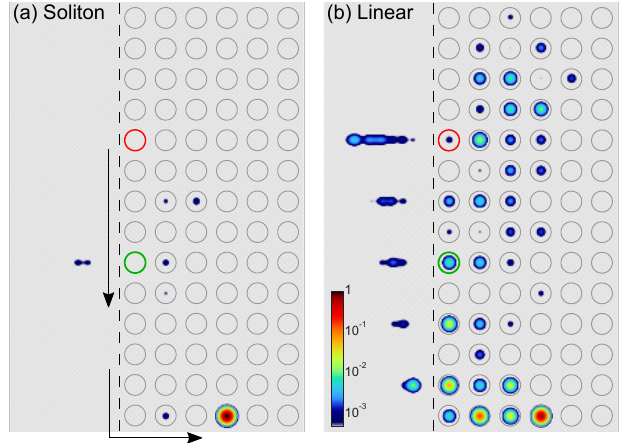}
  \caption{{\color{SMblue}(a) Propagation of single-site traveling soliton along the system bath boundary. A single defect site with a detuned on-site energy $\delta/\Omega\!=\!0.1$ is marked by the green circle. %
  After traveling through the defect site and the lower-left lattice corner, the transmission at the brightest site is $98.6 \%$ and loss to the bath is $0.4\%$. (b) Corresponding linear propagation with increased bath loss ($4.7\%$) and reduced localization of the output state in the lattice. %
  Here, $\Omega=2\pi/Z$, and $Z=26$~mm.
  } 
  }
  \label{supp_defect_corner}
\end{figure}

{\color{SMblue}
\section{Robustness of soliton propagation}
\label{defects}

We have discussed the conditions for forming a traveling edge soliton in a disorder-free, clean lattice coupled to a bath (Section~\ref{optimal-gp-derivation}). In this section, we examine the robustness of the soliton in the presence of defects and sharp corners. We find that the unidirectional propagation of the soliton is maintained, reflecting the topological nature of the underlying linear system. %
In particular, {\it small} defects do not induce 
back-scattering from the soliton. 
However, these imperfections can excite bulk modes and increase bath loss to some extent through nonlinear mode mixing.
Nevertheless, compared with linear dynamics in the same bath-coupled lattice, the soliton exhibits reduced bath loss and stronger, more localized transmission. In other words, the soliton outperforms linear edge transport in our bath-coupled topological system.
Unless otherwise specified, the system parameters used here are the same as those in Fig.~\ref{fig_2} and Fig.~\ref{supp_fig_D}.}

\begin{figure}[t]
  \centering
\includegraphics[width=1\linewidth]{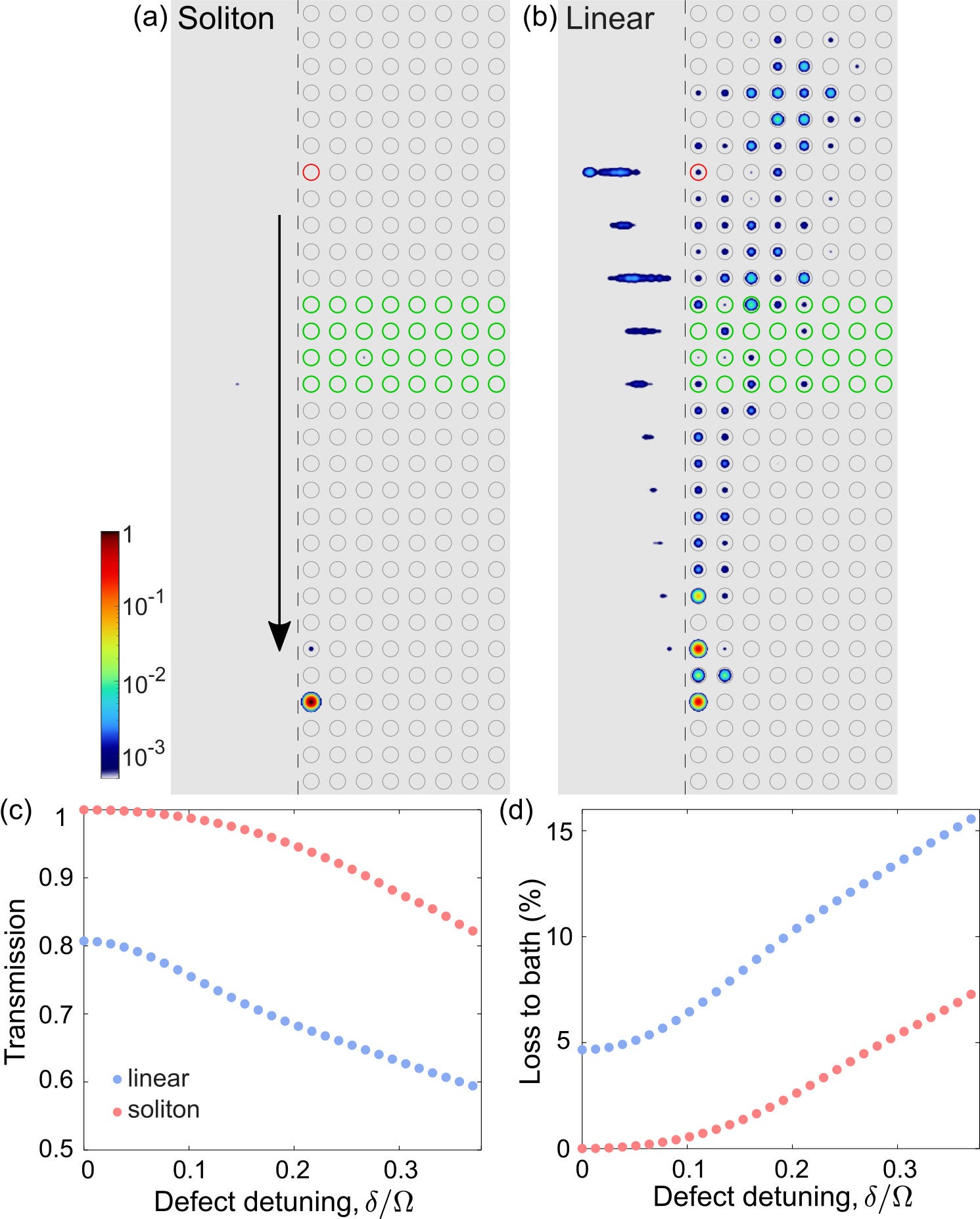}
  \caption{(a) {\color{SMblue}Soliton propagation through a
region of detuned onsite propagation constant, shown by the green circles. Here, $\delta/\Omega=0.1$, where $\Omega=2\pi/Z$, and $Z=26$~mm. The bath loss is found to be $0.5\%$. (b) The corresponding linear dynamics. (c,d) Variation of transmitted power through the defect region and bath loss as a function of on-site detuning $\delta$. The soliton propagation outperforms the linear case in terms of the loss to the bath. Additionally, a higher and more localized transmission is achieved for the soliton.}  
  }
  \label{supp_defect_sweep}
\end{figure}

{\color{SMblue} 
\subsection{A single defect and a sharp corner}

We first consider a bath-coupled topological system with a single defect on the system-bath boundary. Figure~\ref{supp_defect_corner} shows the intensity distribution after $z/Z\!=\!6$ driving periods. The defect site, marked by the green circle, has a detuned on-site energy $\delta$ relative to the rest of the sites. For $\delta\!=\!0$, the soliton propagates unidirectionally around the bottom-left corner without any diffraction into the bulk and loss to the bath (not shown in the figure). As shown in Fig.~\ref{supp_defect_corner}(a), the unidirectional propagation of the state is maintained for $\delta/\Omega=0.1$, and $98.6\%$ of the input optical power is transmitted from the most intense site at the bottom edge. Importantly, no measurable back-scattering is observed, and the loss to the bath is only $0.4\%$. This is an improvement over the linear bath-coupled system [Fig.~\ref{supp_defect_corner}(b)] for which the loss to the bath is higher and the output state in the lattice is not strongly localized. 
}

\begin{figure}[t]
  \centering
\includegraphics[width=0.95\linewidth]{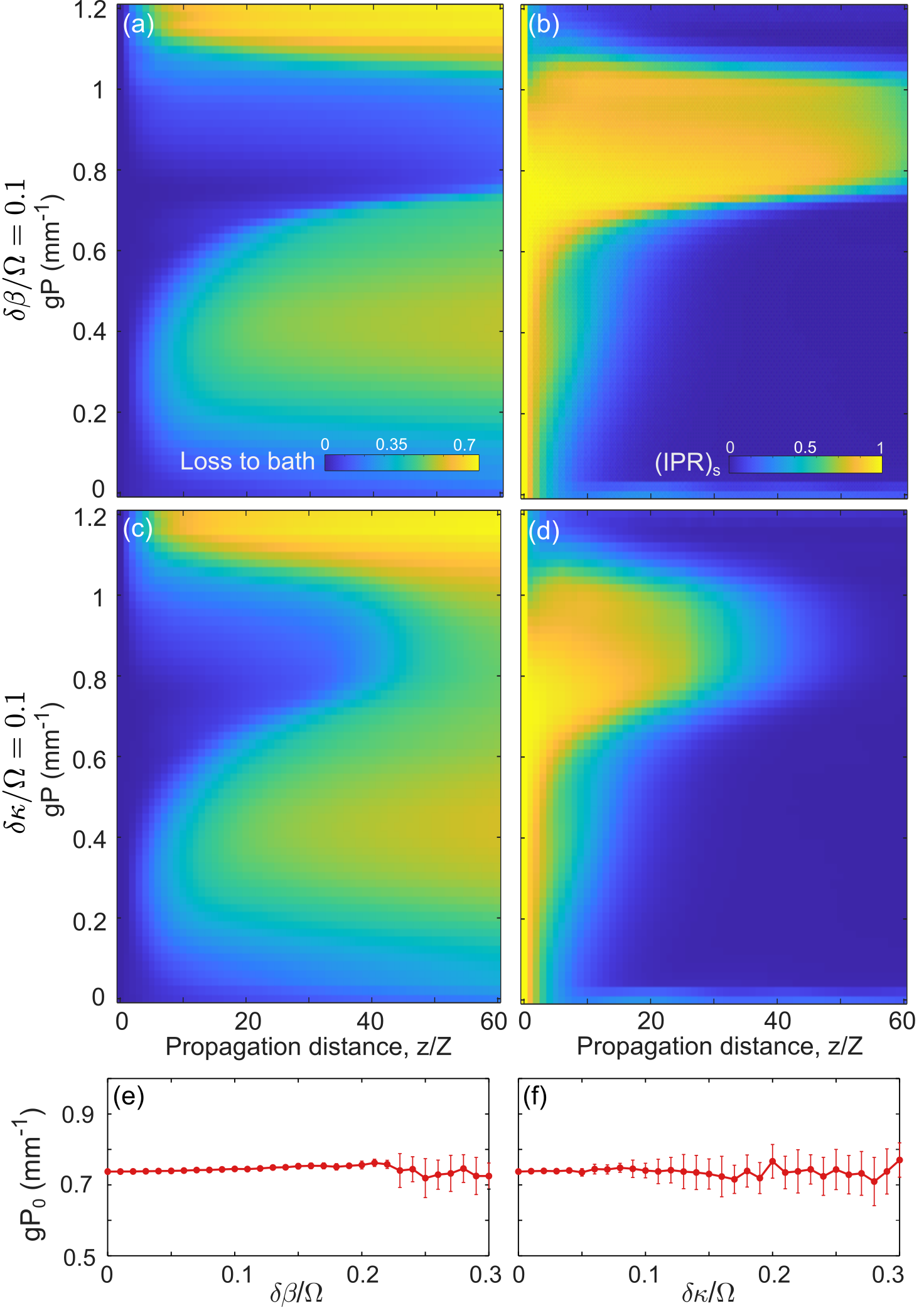}
  \caption{{\color{SMblue}   
  The variation of bath loss and the $\mathrm{(IPR)_s}$ in the presence of random on-site disorder $\delta \beta$ (a,b) and coupling disorder $\delta \kappa$ (c,d). Here, $\Omega=2\pi/Z$, and $Z=26$~mm. In both cases, the soliton is long-lived with a finite lifetime. Nonlinearity-induced bath-loss suppression persists in (a,c) up to a long $z/Z$. 
  Results for the clean case, $\delta \beta\!=\!0\!=\!\delta \kappa$, are presented in Fig.~\ref{supp_fig_D}(a,b). (e,f) The variation of the optimal power associated with the minimum loss as a function of disorder strengths for $z/Z=10$. The soliton tolerates relatively stronger
disorder in propagation constants than in couplings.
All results are averaged over 15 disorder realizations.}
  }
  \label{random_disorder}
\end{figure}

\begin{figure*}[t!] 
  \centering
  \includegraphics[width=0.95\textwidth]{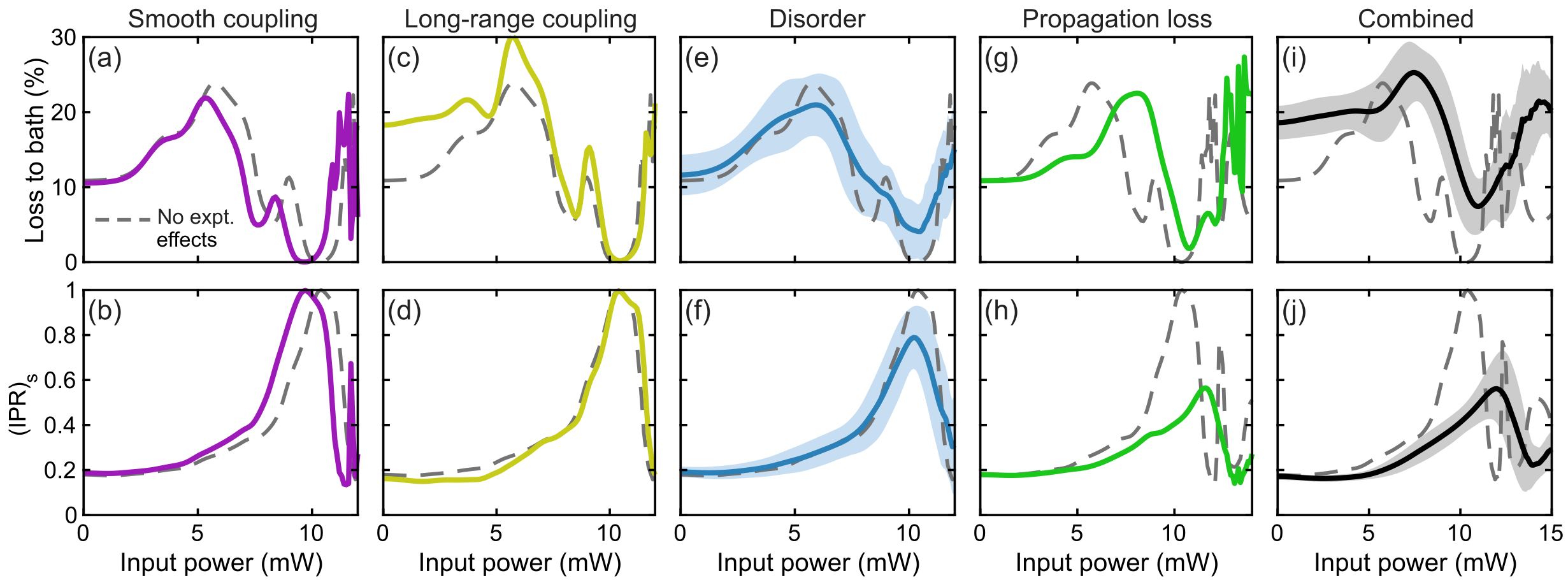}
  \caption{{\color{SMblue}Numerically obtained variation of bath loss (top row) and $(\mathrm{IPR})_s$ (bottom row) as a function of average input power for the bath-coupled topological system. 
  The first four columns individually include the experimental effects mentioned above. In the last column, all four effects are combined.
  The gray dashed line in each subfigure shows the variation in a clean device, without including the experimental effects. Here, $Z=32.8~\mathrm{mm}$, $z/Z\!=\!2$, $\xi_{1,3}=1.40\pi/2$, $\xi_{2,4}=0.98 \pi/2$, and nonlinear coefficient $g=0.091$~mm$^{-1}$ mW$^{-1}$. See text for other parameters.}
  {\color{SMblue}The shaded bands in (e,f) and (i,j) indicate $\pm1$ s.d.\ of $200$ disorder realizations with $\delta\kappa=\delta\beta=0.01$~mm$^{-1}$. The range of the horizontal axis is larger for the last two columns.}
}
  \label{exp_vs_theory_lattice_2}
\end{figure*}

{\color{SMblue} 
\subsection{A strip of defect region%
}

We now consider the soliton propagation through a region of detuned onsite energy, as shown in Fig.~\ref{supp_defect_sweep}. All sites marked by the green circle have an increased onsite energy of $\delta$. Figures~\ref{supp_defect_sweep} (a) and (b) show the intensity distributions after $z/Z\!=\!10$ driving periods for the nonlinear ($gP\!=\!gP_o\!=\!0.74$~mm$^{-1}$) and linear case, respectively. Figs.~\ref{supp_defect_sweep}(c,d) show the variation of power transmitted through the defect region and loss to the bath as a function of on-site detuning. Evidently, no back-scattering is observed for the soliton. Additionally, compared to the linear dynamics, the nonlinear state propagates with lower loss and higher transmission. 
}

{\color{SMblue} 
\subsection{Random disorder}

Finally, we consider a bath-coupled topological system with random disorder in the system. 
The effects of on-site and coupling disorder are investigated separately.
In Figs.~\ref{random_disorder}(a,b), we show the variation of the bath loss and $\mathrm{(IPR)_s}$ in the presence of on-site disorder $\delta \beta$.  On-site propagation constants of the system sites are drawn from a uniform distribution over the range $[-\delta \beta/2, \delta \beta/2]$, and the results are averaged over $15$ disorder realizations. Figures~\ref{random_disorder}(c,d) show similar results in the presence of coupling disorder $\delta\kappa$. In a disorder-free clean lattice, a true traveling soliton is formed at the system-bath boundary with infinite lifetime.
Disorder reduces the lifetime of the soliton. Importantly, nonlinearity-induced dissipation inhibition persists in Figs.~\ref{random_disorder}(a,c).
Up to a long propagation distance, a peak in the $\mathrm{(IPR)_s}$ and a dip in the bath loss can be observed as a function of the nonlinear strength. However, the bath loss is no longer completely suppressed,  as evident in our experimental
observation, see also Section~\ref{Expt_disorder}.

We then consider a bath-coupled system with $z/Z=10$ and examine the variation of the optimal power associated with minimum loss. Figures~\ref{random_disorder}(e,f) show that the optimal power exhibits only minor variations at experimentally relevant disorder strengths. It is evident from Fig.~\ref{random_disorder} that the soliton tolerates relatively stronger disorder in propagation constants than in couplings.
}

{\color{SMblue}
\section{Accounting for experimental effects}
\label{comparison_theory_expt}

As mentioned in the main text, our experiments involve certain effects that cause some deviations of $(\mathrm{IPR})_s$ and bath loss from the behavior shown in Fig.~\ref{fig_2}.
These effects include (i) step-like coupling profiles instead of ideal step functions, (ii) long-range couplings, (iii) moderate disorder in the on-site propagation constant and couplings, and (iv) intrinsic propagation loss.
In this section, we include these effects into our model individually and then collectively to assess their impact on the quantitative agreement between theory and experiment. }

{\color{SMblue} 
We model the fabricated lattice shown in Fig.~\ref{fig_mode-mixing} which consists of a $6\times6$ topological system with three bath chains attached to the boundary A sites on the left.
Each bath chain contains $16$ sites, and the device has a driving period of $Z=32.8$~mm. 
The initial state is localized at the third A site at the system-bath boundary, as indicated by the red circle in Fig.~\ref{fig_mode-mixing}(c), and the output state after two driving periods $z/Z\!=\!2$ is calculated as a function of the nonlinear strength. %
For comparison, Fig.~\ref{exp_vs_theory_lattice_2} shows the bath loss (top row) and $(\mathrm{IPR})_s$ (bottom row) for the clean system alongside the results obtained when experimental effects are included.
In the linear regime, for the clean case without experimental effects, the $(\mathrm{IPR})_s$ is approximately $0.18$ and the bath loss is near $10.9\%$. As the nonlinear strength is increased, a clear peak in $(\mathrm{IPR})_s$, accompanied by a dip in the bath loss that reaches zero, is observed near an average input power of $10.4$~mW.}
%

{\color{SMblue}
\subsection{Step-like coupling profile}\label{Expt_step_like}

In the fabricated lattice, the coupling is switched on and off through waveguide bends with a $\sin^2$-shaped spatial profile over a length of $2$~mm, rather than through ideal step transitions.
When these smooth step-like coupling profiles are considered, while keeping the integrated coupling angles $\xi_m$ unchanged, the results change only marginally [Figs.~\ref{exp_vs_theory_lattice_2}(a,b)]. 
The bath loss still vanishes, and the $(\mathrm{IPR})_s$ clearly shows a peak. 
However, the optimal input power for the complete loss suppression and soliton formation is reduced from $10.4$ to $9.7$~mW. %
This reduction in the optimal power results from the reduced coupling strengths during both the rising and falling segments of the step-like modulation. 
Evidently, the step-like coupling profile is a matter of design choice rather than a fundamental limitation on system performance.}

{\color{SMblue}
\subsection{Long-range couplings}

Outside the second driving step, the bath chains remain at a fixed distance of $29~\mu\mathrm{m}$ from the nearest boundary A sites. Consequently, the system-bath coupling does not vanish completely. 
In other words, in our experiments, the system–bath coupling is given by $\kappa_{sb}(z)+\tilde{\kappa}$, where $\kappa_{sb}(z)$ is active only during the second quarter of each driving period, while $\tilde{\kappa}$ remains constant throughout the driving period.
Based on the $\kappa$-versus-$d$ calibration, we estimate the residual system-bath coupling to be  %
$\tilde{\kappa}=0.017$~mm$^{-1}$ and $\kappa_{sb}=0.12$~mm$^{-1}$. 
This always-on coupling primarily raises the bath loss in the linear regime from $10.9\%$ to $18.3\%$,
because more light is transferred through A sites to the bath [Fig.~\ref{exp_vs_theory_lattice_2}(c)]. 
Importantly, the Kerr nonlinearity can still completely suppress the bath loss, while the variation of  $(\mathrm{IPR})_s$ remains nearly unchanged,
as shown in Fig.~\ref{exp_vs_theory_lattice_2}(d).
The next-nearest neighbor couplings in the topological lattice are estimated to be $0.004$~mm$^{-1}$, which is smaller than the coupling disorder. This weak coupling is included in the model, however, its influence on the present analysis is negligible compared with those of the other four effects. %
}

{\color{SMblue}
\subsection{Moderate disorder}\label{Expt_disorder}

The effect of defects and disorder on soliton propagation has been discussed in Sec.~\ref{defects} at a fixed optimal input power. 
Here, we consider $200$ realizations of random fluctuations in couplings and on-site propagation constants, with %
standard deviations of $\delta\kappa=\delta\beta=0.01$~mm$^{-1}$, to estimate how disorder affects the dependence %
of bath loss and $(\mathrm{IPR})_s$ on the input power relative to the clean case. 
At low input powers, the bath loss varies across disorder realizations as indicated by the shaded region in Figs.~\ref{exp_vs_theory_lattice_2}(e,f). However, the mean bath loss remains close to that of the clean case. 
Disorder makes the minimum of the mean bath loss approximately $4\%$ shallower %
and reduces the maximum of the mean $(\mathrm{IPR})_s$ to approximately $0.73$.} %

{\color{SMblue}We observe a minimum in the bath loss for every disorder realization. %
Moreover, for the disorder strength considered here, the input power at which the loss minimum occurs deviates only slightly from its value in the clean case; see also Figs.~\ref{random_disorder}(e,f). 
This feature is advantageous for photonic applications because the system can operate at reduced bath loss for a given fabrication-induced disorder realization.}

{\color{SMblue}
\subsection{Intrinsic waveguide loss}

The effect of intrinsic waveguide loss is examined in Figs.~\ref{exp_vs_theory_lattice_2}(g,h). 
A clear peak in $(\mathrm{IPR})_s$ and a dip in bath loss are visible. 
However, the peak height of $(\mathrm{IPR})_s$ is reduced to $\approx0.52$, while the minimum bath loss reaches approximately $1.6\%$. Furthermore, because the optical power decays during propagation, both features shift to a slightly higher input power of approximately $11.2$~mW.
These results indicate that intrinsic waveguide loss can limit localization in the lattice. Nevertheless, the loss level in our experiments does not wash out the key features investigated in this work.
Note that the total propagation loss and the total bend loss are estimated to be $3.4$~dB and $1.7$~dB for the device in Fig.~\ref{fig_mode-mixing}. For simplicity, here we consider an overall loss of $5.1$~dB over the entire device length, corresponding to a loss coefficient of $\alpha=0.0154$~mm$^{-1}$.}

{\color{SMblue}
\subsection{Combined effects}

When all the above-mentioned effects are incorporated into the model [Fig.~\ref{exp_vs_theory_lattice_2}(i,j)], the calculated bath loss curve reproduces the main features of the experimental data shown in Figs.~\ref{fig_mode-mixing}(a,b). Specifically, the model predicts a bath loss of approximately $19\%$ in the linear regime, consistent with the measured value of $18.8\%$, followed by a maximum of approximately $24.5\%$ (at $7.6$~mW) at intermediate input powers, compared with the measured $23\%$ over $6.5$--$7.5$~mW. 
At higher input powers, the calculated loss decreases to approximately %
$8\%$ near $11$~mW, which is in good agreement with the measured value of $9.1\%$. 
The calculated $(\mathrm{IPR})_s$ increases from $0.17$ to  $0.49\pm 0.11$, whereas the corresponding experimental values increase from $0.16$ to $0.27$.

These results indicate that a step-like coupling profile is a design consideration rather than a fundamental bottleneck. Moreover, although always-on system–bath coupling increases the linear loss, its effect on dissipation is almost completely suppressed by the nonlinearity. In some applications, this increased linear loss may even be advantageous. Finally, propagation loss primarily limits localization within the lattice, whereas disorder limits the extent of bath-loss suppression. Overall, these results highlight the potential of nonlinearity-induced dissipation suppression for practical photonic devices, where propagation loss and fabrication disorder can be accounted for as distinct design constraints.
}

\section{Multi-site solitons}\label{multisite}

\begin{figure}[h]
  \centering
  \includegraphics[width=1\linewidth]{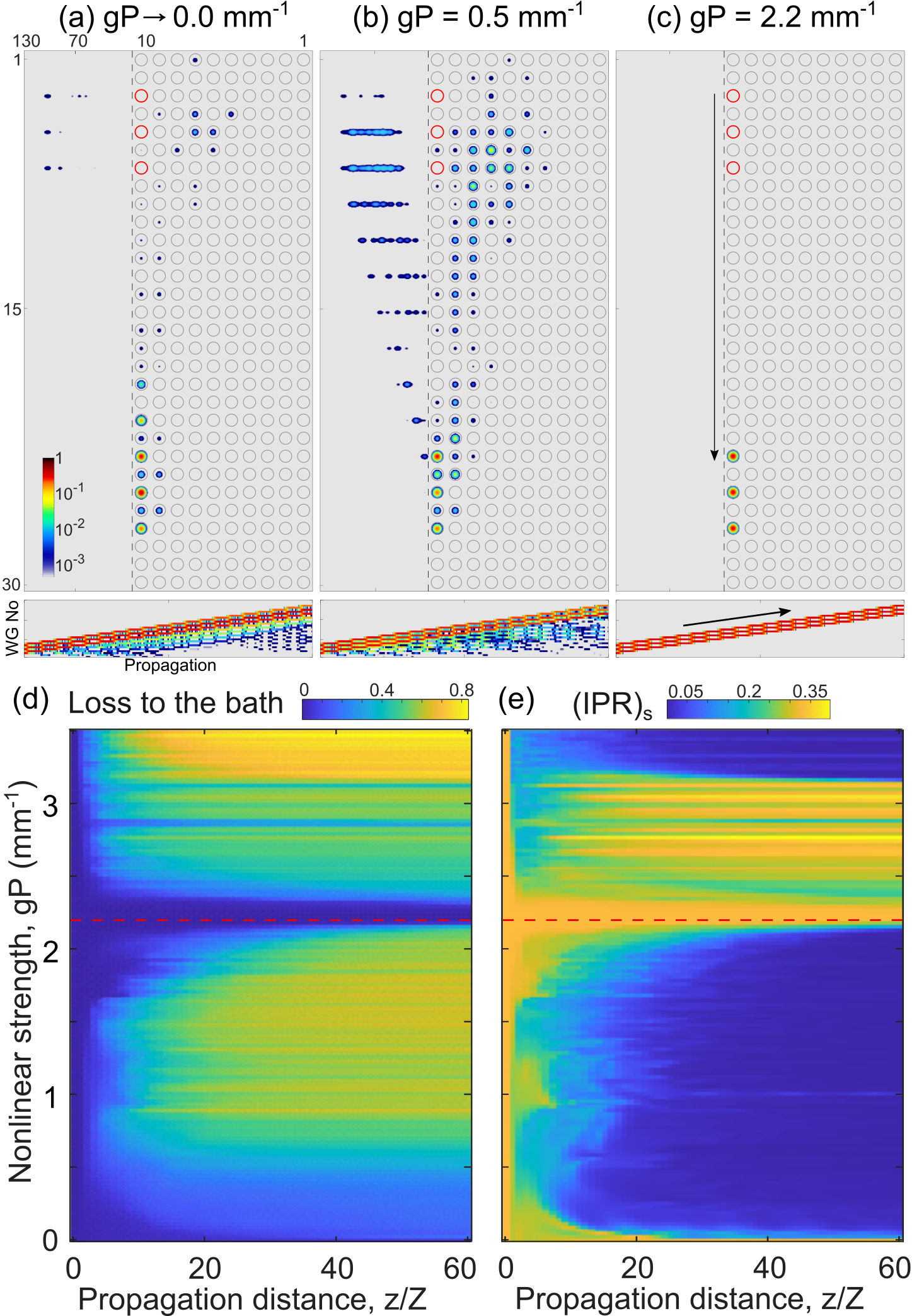}
  \caption{(a–c) Intensity distributions at $z/Z\!=\!10$ for three different nonlinear strengths $gP$. %
  The initial state consists of equal amplitudes on the three boundary sites $3$, $5$, and $7$, marked by red circles. (d,e) Numerically obtained loss to the bath and (IPR)$_s$, respectively. The red dashed line in (d) and (e) indicates the nonlinearity strength %
  for which the three-site traveling soliton forms at the system-bath boundary.
}
  \label{S6}
\end{figure}

In this section, we study how Kerr nonlinearity influences multi-site edge excitations. We use the same system size and coupling parameters as in %
Section~\ref{method-numerics}. Unlike the single A-site excitation in 
Fig.~\ref{fig_2} (main text), here we consider an input state at three consecutive A sites  
along the system-bath boundary with equal amplitudes. 
Figures~\ref{S6}(a–c) show the intensity distributions at
$z/Z\!=\!10$ for three different nonlinear strengths.
The variation of loss to the bath and  $\mathrm{(IPR)_s}$ as a function of $gP$ and propagation distance are shown in Figs.~\ref{S6}(d,e). %
At intermediate nonlinearity, we observe enhanced dissipation due to nonlinear mode-mixing, Fig.~\ref{S6}(b).
In contrast, at an optimal nonlinearity $gP$ %
(marked by the red dashed line), a three-site traveling soliton is formed, as shown in Fig.~\ref{S6}(c). This state travels along the edge %
without any dissipation to the bath
and remains localized.
Note that the nonlinearity needed to form this three-site soliton is %
three times larger than that for the single-site soliton. %

\begin{figure}[]
  \centering
  \includegraphics[width=1\linewidth]{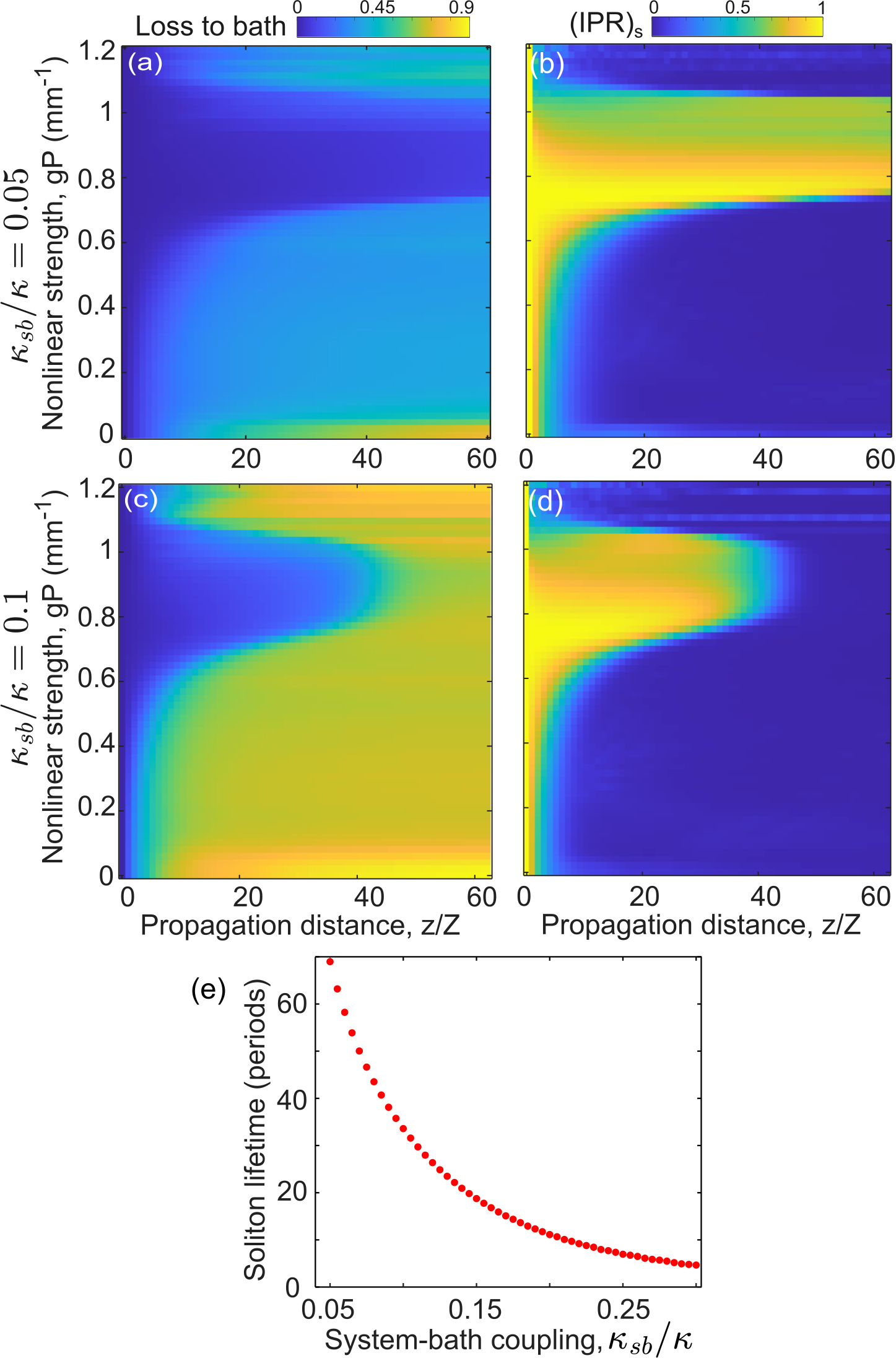}
  \caption{{\color{SMblue}
  Always-on system-bath coupling: Both sublattices (A and B) are coupled to their respective bath chain during all four driving steps. The variations of bath loss  and $\mathrm{(IPR)_s}$ as functions of nonlinear strength $gP$ and propagation distance $z/Z$ for $\kappa_{sb}/\kappa\!=\!0.05$ (a,b) and $0.1$ (c,d). Here, $\xi\!=\!1.15\pi/2$, $\kappa=0.278$~mm$^{-1}$, and the light is launched at a single A site at the system-bath boundary. (e) Soliton lifetime as a function of $\kappa_{sb}/\kappa$. The same system size has been used as in Fig.~\ref{fig_2} in the main text.}
}
  \label{supp_alwayson}
\end{figure}

{\color{SMblue}
\section{
Always-on system-bath coupling 
}

\label{alwayson}
So far, we have considered a specific system-bath coupling -- namely, the bath chains are coupled to the A sites only during the second quarter of each driving period. In this section, we relax these conditions and investigate whether our results can be generalized. Specifically, we now consider that the bath chains are coupled to both A and B sites of the lattice edge during all four driving steps. For this model, we numerically calculate bath loss and $\mathrm{(IPR)_s}$ as functions of $z/Z$ and $gP$.  Figures~\ref{supp_alwayson}(a,b) and (c,d) show these quantities for two different values of the system-bath couplings $\kappa_{\text{sb}}$. In the linear regime, an initial state localized at a single edge A site rapidly loses optical power to the bath. The introduction of an appropriate Kerr nonlinearity localizes the propagating state along the system–bath boundary and substantially suppresses bath-induced dissipation over an extended propagation distance. In other words, in this model, a long-lived traveling edge soliton forms whose lifetime is determined by $\kappa_{\text{sb}}$, as shown in Fig.~\ref{supp_alwayson}(e). Here, we quantify soliton lifetime as the propagation distance where the soliton power drops to $1/e^2$ of the initial power. These results demonstrate that Kerr-nonlinearity-induced dissipation suppression persists over extended propagation distances in this more general bath-coupled model, when the bath is coupled to both sub-lattices throughout the driving period.
An attractive feature of the 
always-on system–bath coupling is that minimum bath loss can be achieved over a range of $gP$ values, rather than at a single $gP$ value, as in the model in Fig.~\ref{fig_1}.
}

\section{Estimation of $\xi$ in experiments}\label{est_lam}
The parameters $\xi_{1, 3}$, $\xi_{2, 4}$, $\kappa_{sb}$, and $\kappa_{\rm b}$ in Figs.~\ref{fig_3} and \ref{fig_mode-mixing}, 
were estimated by comparing numerical and experimental output intensity patterns at the end of the lattice, i.e., at $z/Z  \!=\!  3$, and $2$, respectively. For each case, $6$ to $7$ independent datasets with different input excitations were analyzed. The parameters were varied to achieve the best match of intensities at the system sites (i.e., highest $R$-squared value), and the lowest error in the light lost to the bath. The mean $\xi$ values and their standard deviations ($\delta \xi$) are mentioned in the main text. {\color{SMblue}The $\delta \xi$ values are consistent with the estimated coupling disorder in our experiments. }

\section{Description of supplementary movies}\label{movies}

\noindent
\href{https://sebabrata-mukherjee.github.io/Dissipation_suppression_Topologyn.html}{URL for Supplementary Movie 1-5.}\\

Light propagation in a bath-coupled photonic Floquet topological insulator for {\color{SMblue}five} different nonlinear strengths $gP$, as indicated in each movie.
A single-site initial state (marked by a red circle) is evolved according to Eq.~\eqref{eq1} for ten driving periods. We consider a topological system of size $N_x \times N_y \!=\!40 \times 130$ sites, with each bath chain containing $800$ sites, as in Fig.~\ref{fig_2}. The field of view is smaller than the full system, and the bath is horizontally compressed for presentation purposes. The driving parameter $\xi$ is set to $1.15\pi/2$. The $\mathrm{(IPR)_s}$ and loss to the bath are mentioned at the top of the movies.

\end{document}